\documentclass[showpacs,preprintnumbers,amsmath,amssymb,twocolumn,nofootinbib,nobibnotes,longbibliography,superscriptaddress]{revtex4-1}
\usepackage{amsmath}
\usepackage{amssymb}
\usepackage{bm}
\usepackage{physics}
\usepackage{epsfig}
\usepackage{graphicx}
\usepackage{color}
\usepackage[colorlinks]{hyperref}
\usepackage[T2A]{fontenc}
\usepackage[cp1251]{inputenc}
\usepackage[english]{babel}

\newcommand{\vect}[1]{\boldsymbol{#1}}

\newcommand{\rot}{\mathop{\rm rot}\nolimits}

\usepackage[normalem]{ulem}
\begin{document}

\title{
Control of the exciton valley dynamics in van der Waals heterostructures
}

\author{A.~I.~Prazdnichnykh} 
\affiliation{Ioffe Institute, 26 Polytechnicheskaya, 194021 St.~Petersburg, Russia}
\affiliation{National Research University Higher School of Economics, Saint Petersburg, Russia,}

\author{M.~M.~Glazov}
\affiliation{Ioffe Institute, 26 Polytechnicheskaya, 194021 St.~Petersburg, Russia}

\author{L. Ren}
\author{C. Robert}
\author{B. Urbaszek}
\author{X. Marie}
\affiliation{Universit\'e de Toulouse, INSA-CNRS-UPS, LPCNO, 135 Avenue Rangueil, 31077 Toulouse, France}

\date{\today}

\begin{abstract}
The exciton valley dynamics in van der Waals heterostructures with transition
metal dichalcogenides monolayers is driven by the long-range exchange interaction
between the electron and the hole in the exciton. 
%We develop a theory of exciton valley dynamics in van der Waals heterostructures with transition metal dichalcogenides monolayers. The dynamics is driven by the long-range exchange interaction between the electron and the hole in the exciton. 
It couples the states active in the opposite circular polarizations resulting in the longitudinal-transverse splitting of excitons propagating in the monolayer plane. Here we study theoretically the effect of the dielectric environment on the long-range exchange interaction and demonstrate how the encapsulation in the hexagonal boron nitride modifies the exciton longitudinal-transverse splitting. We calculate the exciton spin/valley polarization relaxation due to the long-range exchange interaction and demonstrate that the variation of the monolayer environment results in significant, up to five-fold, enhancement of the exciton valley polarization lifetime.
\end{abstract}

\maketitle

\section{Introduction}

Two-dimensional (2D) materials combined into van der Waals heterostructures offer a versatile platform with unusual optical and transport properties~\cite{Geim:2013aa,Butler:2013a}. In the family of the monolayer semiconductors based on transition metal dichalcogenides the optical properties are controlled by robust excitons, Coulomb bound electron-hole pairs~\cite{Splendiani:2010a,Mak:2010bh,PhysRevLett.113.026803,Chernikov:2014a,Yu30122014,Kolobov2016book,RevModPhys.90.021001,Durnev_2018,Li2020}. Tailoring the environment of the monolayer, e.g., by encapsulation into hexagonal boron nitride (hBN), affects excitonic states~\cite{Raja:2017aa,Raja:2019aa}, optical spectra of atomically-thin semiconductors~\cite{PhysRevMaterials.2.011001}, and makes it possible to control radiative lifetime of excitons~\cite{PhysRevLett.123.067401,PhysRevLett.124.027401,PhysRevResearch.2.012029,Horng:19}. It opens wide prospects for nanophotonic applications~\cite{Baranov:2017aa,PhysRevLett.120.037402,PhysRevLett.120.037401,PhysRevB.98.161113,Krasnok:18}.

The direct optical transitions in transition-metal dichalcogenide monolayers involve the electronic states at the edges of the Brillouin zone. The symmetry and spin-orbit interaction enable the so-called chiral selection rules: The band-edge optical transitions at the $\bm K_+$ ($\bm K_-$) valley are induced by the photon of the $\sigma^+$ ($\sigma^-$) circular polarization, since valley and spin are locked~\cite{Xu:2014cr,Xiao:2012cr,Mak:2012qf,Sallen:2012qf,Kioseoglou,Zeng:2012ys,Jones:2013tg}. Accordingly, the optically active excitons possess a valley or pseudospin degree of freedom~\cite{Yu30122014,Yu:2014fk-1,RevModPhys.90.021001}.

The valley dynamics of excitons in transition metal dichalcogenides monolayers is in focus of the experimental and theoretical research nowadays~\cite{PhysRevResearch.1.022007,Shinokita:2019aa,Tang:2019ab,Dufferwiel:2017aa,PhysRevB.101.115307}. It has been established~\cite{glazov2014exciton,PSSB:PSSB201552211,Yu:2014fk-1,PhysRevB.93.045414} that similarly to the case of conventional quasi-two-dimensional semiconductors where the pseudospin is associated with the spins of electron and hole forming an exciton~\cite{maialle93,goupalov98,ivchenko05a}, the bright exciton valley dynamics is controlled by the long-range exchange interaction between the electron and the hole. The process of valley depolarization of the exciton can be considered as a virtual recombination of the electron-hole pair in one valley and its emergence in the opposite valley~\cite{glazov2014exciton,PSSB:PSSB201552211}.

Since the radiative properties of the excitons can be manipulated in van der Waals heterostructures~\cite{PhysRevLett.123.067401,PhysRevLett.124.027401,PhysRevResearch.2.012029}, it is natural to ask the question whether the dielectric environment affects the valley dynamics of excitons in two-dimensional semiconductors. Here we address this question theoretically. We demonstrate that the presence of surrounding hBN layers screens the long-range exchange interaction and slows-down valley depolarization of excitons. We develop a microscopic theory of the effect based on the electrodynamical approach for calculating the exchange interaction in the exciton. We use the density matrix method to study the valley polarization dynamics in transition metal dichalcogenide monolayers. We demonstrate significant, up to a five-fold, increase of the valley polarization lifetime in van der Waals heterostructures depending on the hBN layer thickness. To the best of our knowledge, this control of the exciton spin dynamics by the environment was never demonstrated before in semiconductors. 

The paper is organized as follows: Section~\ref{sec:fine} presents the calculations of the exciton fine structure due to the long-range exchange interaction in van der Waals heterostructures. Next, in Sec.~\ref{sec:dynamics} the valley dynamics of the excitons is calculated and analyzed. Various regimes of valley polarization decoherence depending on the systems' parameters are identified and analyzed. The concluding remarks are presented in Sec.~\ref{sec:concl}.

\section{Exciton fine structure}\label{sec:fine}

This section presents the microscopic theory of the exciton fine structure induced by the long-range exchange interaction between the electron and the hole. The long-range exchange interaction is the driving force for the pseudospin or valley dynamics of excitons in semiconductors~\cite{BP_exch71,denisovmakarov,birpikus_eng,maialle93,ivchenko05a,glazov2014exciton}. It can be calculated either quantum-mechanically by evaluating the matrix elements of the Coulomb potential over properly symmetrized two-particle Bloch functions or electrodynamically, taking into account the self-consistent action of the electric field induced by the exciton. The equivalence of these approaches has been established for two-dimensional semiconductors in Ref.~\cite{glazov2014exciton}. The electrodynamical approach has an advantage of being easily adapted for treatment of inhomogeneous structures such as the one studied here. Thus, we resort to the electrodynamical approach in this work. 

\begin{figure}[htb]
	\centering
	\includegraphics[width=0.9\linewidth]{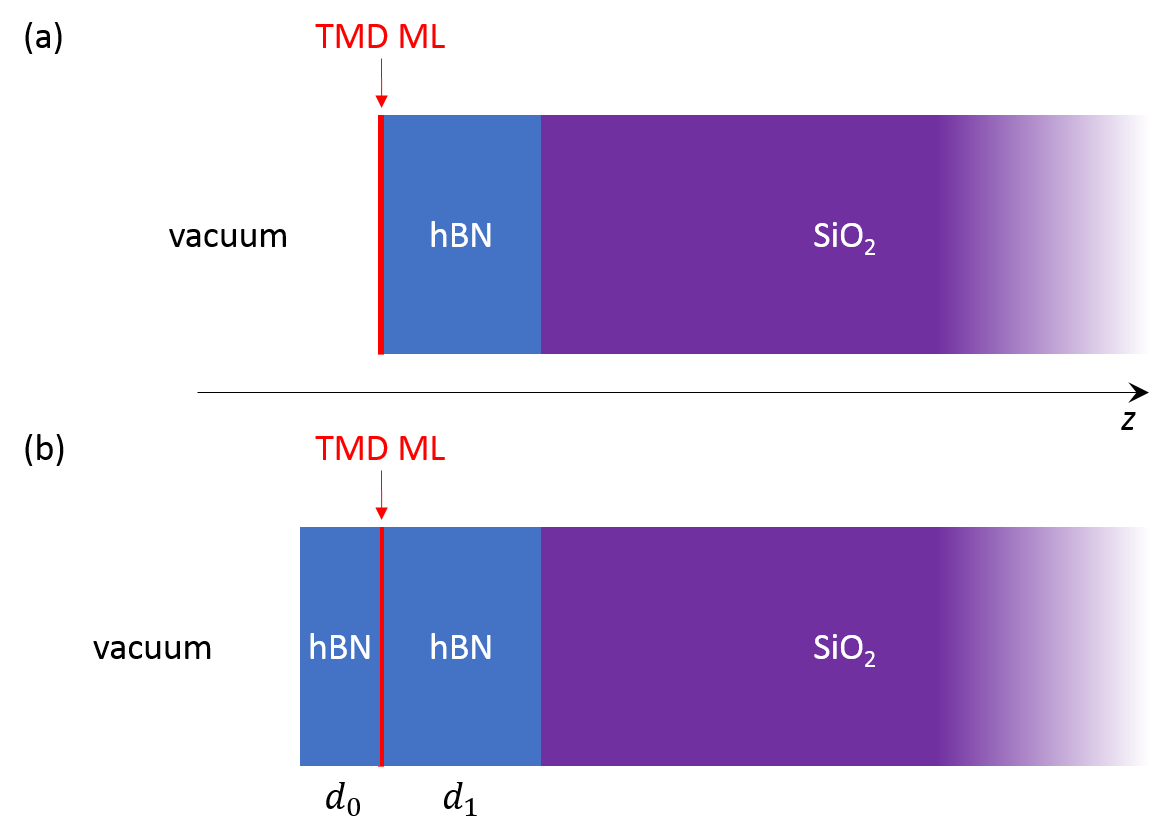}
	\caption{\textbf{Schematics of investigated van-der-Waals heterostructures.}		(a) Structure without a cap hBN layer: TMD monolayer\textendash hexagonal boron nitride (hBN)\textendash silicon dioxide ($\rm{SiO}_2$); (b) Structure with a cap hBN layer: hBN\textendash TMD monolayer\textendash bottom hBN layer\textendash $\rm{SiO}_2$.}
	\label{ris:stuctures}
\end{figure}

\subsection{Structure and modes of electromagnetic field}

We consider the van der Waals heterostructures based on transition metal dichalcogenides (TMD) monolayers (MLs) schematically depicted in Fig.~\ref{ris:stuctures}. This corresponds to the typical stacking of the the encapuslated TMD monolayers investigated in most of the experiments~\cite{PhysRevX.7.021026,Ajayi_2017}.
Two types of structures are analysed: without and with cap hBN layer shown in Fig.~\ref{ris:stuctures} (a) and (b), respectively.

Within the framework of the electrodynamical approach the optically active exciton in TMD monolayer is considered as an in-plane electric dipole or current, oscillating at the optical transition frequency $\omega_0$, where
\begin{equation}
\hbar\omega_0 = E_g - E_B + \frac{\hbar^2 K^2}{2m}. 
\end{equation}
Here $E_g$ is the band gap of the monolayer, $E_B$ is the exciton binding energy, $\bm K$ is the in-plane wavevector of the 2D exciton and $m$ is its effective mass. In the relevant range of the wavevectors the dependence of $\omega_0$ on $\bm K$ can be disregarded. We also note that the variation of the environment, i.e., the thicknesses of the hBN layers, does not strongly affect the difference $E_g-E_B$, the spectral position of the exciton, while both band gap and exciton binding energy strongly depend on the screening~\cite{PhysRevB.97.041409,PhysRevB.98.245126}. The induced current density can be written as~\cite{glazov2014exciton}
\begin{equation}
\label{current:dens}
\bm j(z) = \bm J \delta(z), \quad \bm J = \frac{c\omega_0}{2\pi\omega} \frac{i\Gamma_0}{\omega -\omega_0 + i \Gamma} \bm E_\parallel(z), 
\end{equation}
where $\bm J$ is the 2D (surface) current density, $\Gamma_0$ and $\Gamma$ are the radiative (into vacuum) and non-radiative decay rates of the exciton, the monolayer is assumed to be in the $z=0$ plane, subscript parallel ($\parallel$) denotes the in-plane components of the field. In Eq.~\eqref{current:dens} $\bm E$ is the electric field acting on the exciton, which includes both the external field an the field induced by the exciton, $\omega$ is the frequency of the field. The thickness of the monolayer is negligble as compared with the wavelength of light emitted by the exciton, that is why it is sufficient to use strictly two-dimensional model for the current density, Eq.~\eqref{current:dens}.

In what follows we apply the uniaxial approximation for description of the excitonic states. In such a case the eigenmodes of the system, being the coupled modes of the exciton and electromagnetic field, can be described by the wavevector $\bm K$ in the monolayer plane and polarization: the $s$-polarization corresponds to the $\bm j(z)\perp \bm K_\parallel$, this state is denoted as the transversal (T) exciton and the $p$-polarization corresponds to the $\bm j(z)\parallel \bm K_\parallel$, this state is denoted as the longitudinal (L) exciton.

\begin{figure*}[htb]
    \centering
	\includegraphics[width=0.9\linewidth]{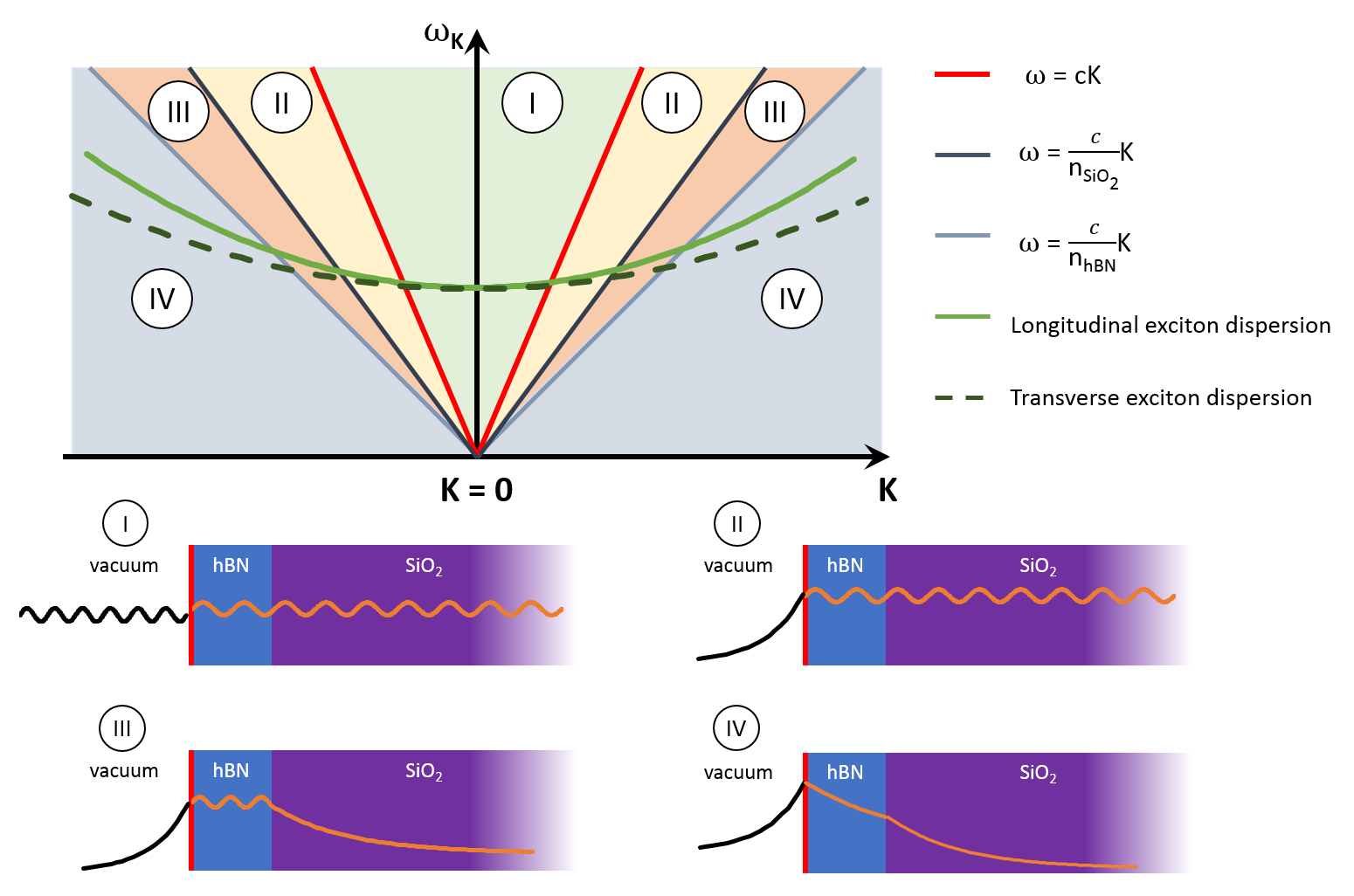}\\
	\caption{\textbf{Exciton dispersion and induced electromagnetic field.} Top panel: schematic illustration of longitudinal (L, solid curve) and transverse (T, dashed line) exciton energy spectra. $\textbf{K} = 0$ point corresponds to center of the exciton Brillouin zone. Dispersion is shown not to scale. Four bottom panels (I) \ldots (IV) show schematics of the electromagnetic field distribution depending on area where the exciton wavevector lies. Wavy lines correspond to propagating waves
	 $\propto \exp(ik_z\abs{z})$ induced by the exciton, decreasing curves correspond to decaying waves $\propto \exp(-\varkappa_z\abs{z})$.}
	\label{ris:light_cone}
\end{figure*}

The current $\bm j(z)$ in Eq.~\eqref{current:dens} is associated with the exciton induced electromagnetic field, which can propagate away from the monolayer or decay with the distance from the monolayer depending on the exciton wave vector. These regimes of field propagation are shown in Fig.~\ref{ris:light_cone}. Top panel shows the light cones, i.e., the dispersion of free electromagnetic waves in the vacuum, SiO$_2$, and hBN. In the case where the wavevector of the exciton lies in areas I or II, 
\[
	K \leqslant \frac{\omega_0}{c} n_{\rm{SiO}_2},
\]
the exciton emits propagating waves, which cause its radiative decay. Interestingly, for $K\leqslant \omega_0/c$ the waves are propagating both to the vacuum and to the substrate (region I), while for $\omega_0/c \leqslant K \leqslant \omega_0  n_{\mathrm{SiO}_2}/c$ the field decays into vacuum but propagates into SiO$_2$. A self-consistent interaction of such exciton with the induced field leads to the difference of the decay rates for the longitudinal and transverse excitons~\cite{glazov2014exciton}. 
The exciton with the wavevector outside the SiO$_2$ light cone, areas III and IV in Fig.~\ref{ris:light_cone}, induces exponentially decaying waves (both into the vacuum and substrate). Here, its interaction with the self-consistent field results in the renormalization of the longitudinal and transverse excitons energies. We will mainly focus on the latter case where $K\geqslant\omega_0 n_{\rm{SiO}_2}/c$, as for typical experimental parameters the states outside the light cones are mostly populated. 

In order to find the exciton energy spectrum fine structure we have to self-consistently solve the Maxwell's equations
\begin{subequations}
    \label{eq:maxwell:0}
\begin{align}
    \rot \vect{E} &= -\frac{1}{c}\pdv{\vect{B}}{t},\label{maxwell:E}\\
    \rot \vect{B} &= \frac{1}{c}\pdv{\vect{D}}{t} + \frac{4\pi}{c} \bm j(z),\label{c}
\end{align}
\end{subequations}
together with the Eq.~\eqref{current:dens} for the exciton-induced current and explicit expression for the electric induction $\bm D = \varepsilon(z) \bm E$ with $\varepsilon(z)$ being the high-frequency (background) dielectric constant of the structure found disregarding excitonic effects. In this way, both the damping of the exciton and its energy renormalization due to the long-range exchange interaction can be derived automatically accounting for the screening and retardation effects~\cite{zhilich74:eng,Kiselev74,glazov2014exciton}.
To that end, it is convenient to include the current $\bm j(z)$ into the boundary condition for the in-plane components of magnetic field, namely, 
\begin{equation}
\label{bound:B}
\vect{B}_{\parallel}\left(z \rightarrow 0+\right) - \vect{B}_{\parallel}\left(z \rightarrow 0-\right) = \frac{4\pi}{c}[\bm J\times\vect{e}_z],
\end{equation}
with $\bm e_z$ being the unit vector along the normal to the ML. The remaining boundary conditions are the standard ones implying continuity of the in-plane components of $\bm E$ and normal components of $\bm D$ and $\bm B$ at the interfaces. Below we present the results of solution of the Maxwell's equations and the analysis of the exciton fine structure.

\subsection{Structure without a cap layer}\label{subsec:nocap}

It is instructive to analyze in detail the eigenstates of the exciton coupled with electromagnetic field in the simplest structure without a hBN  cap layer, Fig.~\ref{ris:stuctures}(a). Let us enumerate the layers of the structure: $i = 0$ is the vacuum ($z < 0$), $i = 1$ is the substrate hBN layer ($0 < z < d$) and $i = 2$ is the substrate $\rm{SiO}_2$ ($z > d$). Inside each bulk layer we can write combining Eqs.~\eqref{eq:maxwell:0}:
\begin{equation}
    \label{eq:wave}
    \rot\rot\vect{E} = -\frac{\varepsilon_i}{c^2}\pdv[2]{\vect{E}}{t}, 
%    \Rightarrow - \vect{k}_i\times\left(\vect{k}_i\times\vect{E}\right) = \varepsilon_i\frac{\omega^2}{c^2}\vect{E} \Rightarrow \\ \Rightarrow \left\{\begin{array}{l}
%	\vect{E}\cdot\vect{k}_i = 0\\
%	\omega = \frac{ck_i}{\sqrt{\varepsilon_i}}
%	\end{array}\right.,
\end{equation}
where $\varepsilon_i$ is the dielectric permittivity of the $i$th layer. We seek the solution of Eq.~\eqref{eq:wave} in the form of a plane wave in each layer
\begin{equation}
\label{field}
\bm E(\bm r) = \bm E^{(i)} e^{i \bm k_i \cdot \bm r - i \omega t},
\end{equation}
with $\bm E^{(i)}$ being its complex amplitude, $\vect{k}_i$ is the light wavevector at the frequency of $\omega$ in the $i$th layer. Naturally, we find the absolute value of the wavevector $k_i = \sqrt{\varepsilon_i}\omega/c$. Without loss of generality, we set $\vect{k}_i = \left(0,k_{i,y},k_{i,z}\right)$ and take into account that its $y$-component, $k_{i,y} \equiv k_y$ remains constant in each dielectric layer because of the translational invariance of the system in the $(xy)$ plane. The $z$-component of the wavevector reads $k_{i,z} = \sqrt{\varepsilon_ik_z^2 + (\varepsilon_i - 1)k_y^2}$, with $k_z$ being the wavevector component in the vacuum.

There are two eigenmodes of the electromagnetic field in each layer, namely, TE (or T-) mode and TM (or L-) mode, whose eigenvectors read
\begin{multline}
	\vect{E}_{TE}^{(i)} = \left(\begin{array}{c}
	E_x^{(i)}\\
	0\\
	0
	\end{array} \right),\;~~ \mbox{and}\quad  \vect{E}_{TM}^{(i)} = \left(\begin{array}{c}
	0\\
	E_y^{(i)}\\
	-\frac{k_y}{k_{i,z}}E_y^{(i)}
	\end{array} \right).
\end{multline}
Correspondingly, the TE-mode couples with the transversal exciton and the TM-mode couples with the longitudinal exciton.

Further we have to construct the solution that satisfies the boundary conditions at the interfaces. Let us start with the TE-mode. We seek a solution for the electric field in a form  (we assume the amplitude of the wave at $z\to -0$ wave equals to unity):
\begin{equation}
    \label{eq:TE_el}
    E_x = e^{ik_yy}\left\{\begin{array}{l}
    e^{\varkappa_{0,z}z},\quad z < 0,\\
    E_1e^{ - \varkappa_{1,z}z} + E_2e^{ \varkappa_{1,z}z},\quad 0<z<d,\\
    E_3e^{- \varkappa_{2,z}(z - d)}, \quad z > d.
    \end{array}\right.
\end{equation}
Here we have selected the form of the fields relevant for the states outside the light cone ($k_y > \max\limits_i\sqrt{\varepsilon_i}(\omega/c)$, region IV in Fig.~\ref{ris:light_cone}) where $\varkappa_{i,z} = [{k_y^2 - \varepsilon_i(\omega/c)^2}]^{1/2}>0$, which decay to the both sides of the structure. The boundary conditions of the continuity of the tangential components of the electric field read
\begin{subequations}
    \label{eq:edge_el}
\begin{align}
    &E_1 + E_2 = 1,\\
    &E_3 = E_1e^{-\varkappa_{1,z}d} + E_2e^{\varkappa_{1,z}d}.
\end{align} 
\end{subequations}
Expressing the tangential components of the magnetic field from Eq.~\eqref{maxwell:E} and making use of the boundary condition~\eqref{bound:B} at $z=0$ and the continuity condition at $z=d$ we have
\begin{subequations}
    \label{eq:edge_mag}
%    \\at $z = 0$ (monolayer plane):
    \begin{equation}
    \label{eq:edge_mag_0}
        \varkappa_{1,z}(E_1 - E_2) + \varkappa_{0,z} = \frac{\left(\omega/c\right)^2}{\omega_0/c}\frac{2\Gamma_0}{\omega_0 - \omega - i\Gamma},
    \end{equation}
    \begin{equation}
        \varkappa_{2,z}E_3 = \varkappa_{1,z}E_1e^{-\varkappa_{1z}d} - \varkappa_{1,z}E_2e^{\varkappa_{1,z}d}.
    \end{equation}
\end{subequations}
The boundary conditions Eqs.~\eqref{eq:edge_el} and \eqref{eq:edge_mag} represent the set of four equations for three amplitudes $E_1,E_2,E_3$. Its compatibility condition allows us to find the renormalized energies of excitons due to the light-matter interaction, i.e., taking into account the long-range exchange interaction. As we are looking for the relatively small renormalizations of the exciton energy, where $\abs{\hbar\omega - \hbar\omega_0} \ll \hbar\omega_0$, it is accurate to replace $\omega$ with $\omega_0$ everywhere except for the denominator in the right side of Eq.~\eqref{eq:edge_mag_0}. 
As a result, combining Eqs.~\eqref{eq:edge_el} and \eqref{eq:edge_mag}, we find for the eigenfrequency $\omega_T \equiv \omega$ of the transversal exciton:
\begin{subequations}
\label{nocap:renorm}
\begin{equation}
    \frac{\omega_{T} - \omega_0}{\Gamma_0} = -\frac{2\zeta}{s_0 - s_1 + \frac{2s_1(s_1 + s_2)}{s_1 + s_2 + \left(s_1 - s_2\right)e^{-2as_1} }}.
    \label{eq:s-mod}
\end{equation}
Here the following notations are introduced
\begin{equation}
\label{not:1}
\zeta = \frac{1}{K}\frac{\omega_0}{c}, \quad s_i = \sqrt{1 - \varepsilon_i\zeta^2}, \quad a = Kd.
\end{equation}

Analogous calculation for the TM-polarized mode yields the eigenfrequency $\omega_L$ of the longitudinal exciton:
\begin{multline}
    \frac{\omega_{L} - \omega_0}{\Gamma_0}\\
     = \frac{2}{\frac{\zeta}{s_0} + \frac{\varepsilon_1\zeta}{s_1}\left(\frac{2(\varepsilon_2s_1 + \varepsilon_1s_2)}{\varepsilon_1s_2 + \varepsilon_2s_1 + \left(\varepsilon_1s_2 - \varepsilon_2s_1\right)e^{-2as_1}} - 1\right)}.
    \label{eq:p-mod}
\end{multline}
\end{subequations}

Equations~\eqref{eq:s-mod} and \eqref{eq:p-mod} describe the dispersion of the transveral and longitudinal excitons in the van der Waals heterostructure without a cap layer, Fig.~\ref{ris:stuctures}(a).

\subsection{Effect of the cap hBN layer}\label{subsec:cap}

Now we consider a van der Waals heterostructure capped with a hBN layer as the ones used in most of the experiments, Fig. \ref{ris:stuctures}(b). Explicit expression for exciton energies with account for the exchange interaction can be found by solving the set of Maxwell's equations~\eqref{eq:maxwell:0} with appropriate boundary conditions. As the solution is completely analogous to that presented above in Sec.~\ref{subsec:nocap} for an uncapped structure and quite lengthy, we just give here the results for the radiative doublet eigenfrequencies:
\begin{subequations}
\label{cap:renorm}
    \begin{equation}
        \frac{\omega_{T} - \omega_0}{\Gamma_0} = -\frac{\zeta}{\sqrt{1 - \varepsilon_1\zeta^2}}\frac{\left(1 - r_{1,s}\xi_{0}\right)\left(1 + r_{b,s}\xi_{1}\right)}{1 + r_{1,s}r_{b,s}\xi_{0}\xi_{1}},\label{cap:T}
    \end{equation}
    \begin{equation}
        \frac{\omega_{L} - \omega_0}{\Gamma_0} = \frac{\sqrt{1 - \varepsilon_1\zeta^2}}{\varepsilon_1\zeta}\frac{\left(1 - r_{1,p}\xi_{0}\right)\left(1 + r_{b,p}\xi_{1}\right)}{1 + r_{1,p}r_{b,p}\xi_{0}\xi_{1}}.\label{cap:L}
    \end{equation}
\end{subequations}
Here 
\begin{equation}
\label{not:2}
\xi_i = \exp\left[-2d_i\frac{\omega_0}{c}\frac{\sqrt{1 - \varepsilon_1\zeta^2}}{\zeta}\right],
\end{equation}
 $d_i$ are the thicknesses of the cap, $i = 0$, and the substrate, $i = 1$, hBN layers; $r_{1,\alpha}$ is the reflection coefficient of $\alpha=s,p$-polarized light from vacuum\textendash hBN interface, expressed using Fresnel's equations~\cite{ll8_eng}:
%\begin{subequations}
    \begin{equation}
        r_{1,s} = \frac{1 - \sqrt{\frac{1 - \varepsilon_1\zeta^2}{1 - \zeta^2}}}{1 + \sqrt{\frac{1 - \varepsilon_1\zeta^2}{1 - \zeta^2}}}, \quad
%    \end{equation}
%    \begin{equation}
        r_{1,p} = \frac{\varepsilon_1 - \sqrt{\frac{1 - \varepsilon_1\zeta^2}{1 - \zeta^2}}}{\varepsilon_1 + \sqrt{\frac{1 - \varepsilon_1\zeta^2}{1 - \zeta^2}}}\;;
    \end{equation}
%\end{subequations}
and $r_{b, \alpha}$ is the reflection coefficient from the hBN\textendash $\rm{SiO}_2$ interface:
%\begin{subequations}
    \begin{equation}
        r_{b,s} = \frac{1 - \sqrt{\frac{1 - \varepsilon_2\zeta^2}{1 - \varepsilon_1\zeta^2}}}{1 + \sqrt{\frac{1 - \varepsilon_2\zeta^2}{1 - \varepsilon_1\zeta^2}}}, \quad
%    \end{equation}
%    \begin{equation}
        r_{b,p} = \frac{\frac{\varepsilon_2}{\varepsilon_1} - \sqrt{\frac{1 - \varepsilon_2\zeta^2}{1 - \varepsilon_1\zeta^2}}}{\frac{\varepsilon_2}{\varepsilon_1} + \sqrt{\frac{1 - \varepsilon_2\zeta^2}{1 - \varepsilon_1\zeta^2}}}.
    \end{equation}
%\end{subequations}
As expected, at $d_0=0$ Eqs.~\eqref{cap:renorm} are identical to Eqs.~\eqref{nocap:renorm}.

\begin{figure*}[tb]
\includegraphics[width=\linewidth]{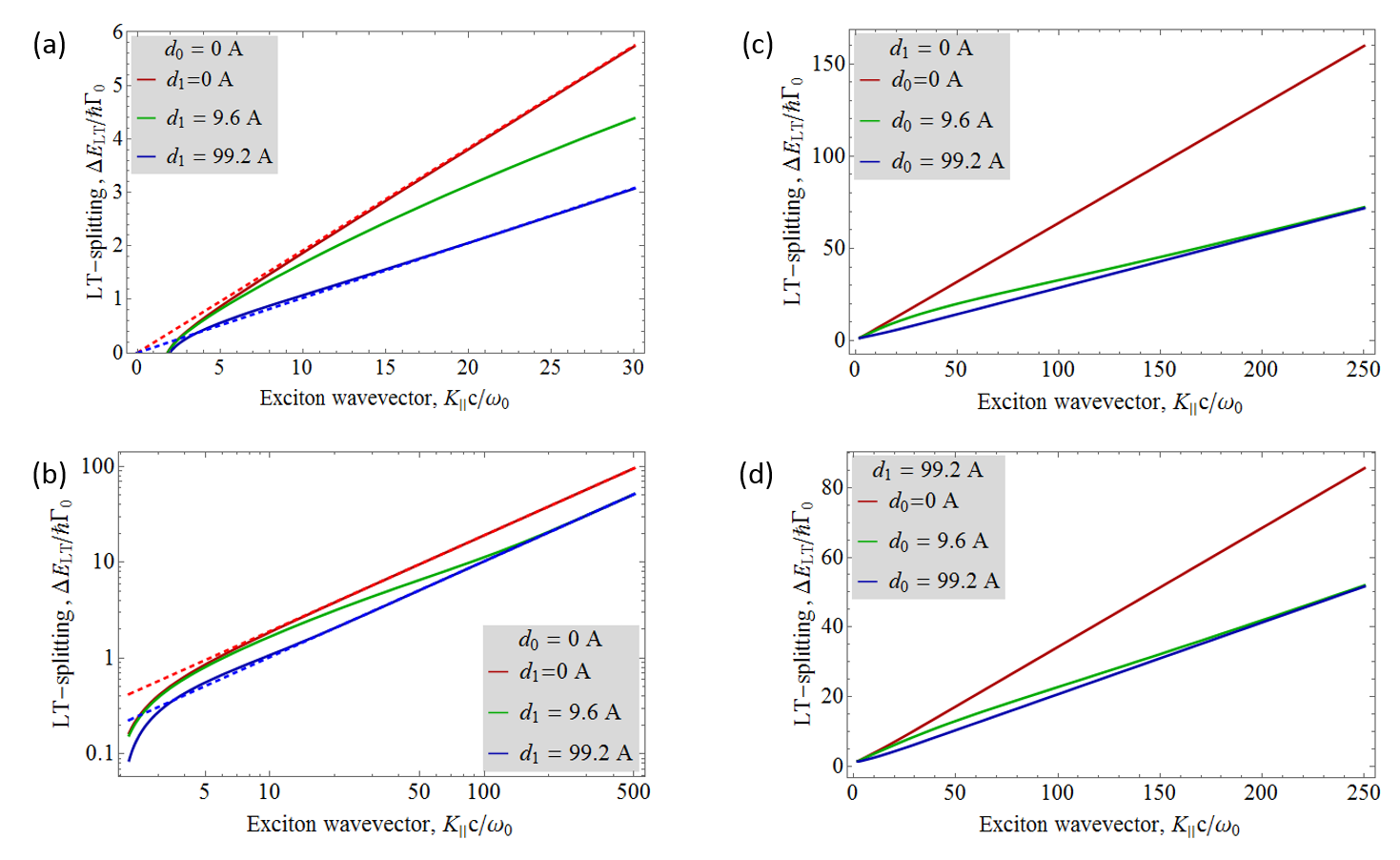}
\caption{\textbf{Exciton fine structure splitting.} LT-splitting as a function of the exciton wavevector in the structure without a cap hBN layer ($d_0=0$) in linear scale (a), log--log scale (b); and in the structures without a substrate hBN layer ($d_1=0$) (c), and with the substrate hBN layer of finite thickness ($d_1=99.2$ \AA) (d), while the cap hBN layer thickness being varied. Dashed lines illustrate asymptotics, Eq.~\eqref{eq:assympt_lt}: blue dashed line corresponds to $d_{\rm{hBN}}\neq 0$ asymptotics, red dashed line corresponds to $d_{\rm{hBN}} = 0$. Parameters of the calculation are: $\varepsilon_{\rm hBN} = 4.84$ is the hBN permittivity, $\varepsilon_{\rm SiO_2} = 2.13$ is the SiO$_2$ permittivity, $\omega_0/c = 981745$ cm$^{-1}$ (which roughly corresponds to MoS$_2$ ML), $\hbar\Gamma_0 = 0.3$ meV.} 
\label{ris:lt_splitting}
\end{figure*}

\subsection{Effect of the hBN layers on the exciton fine structure}
In agreement with the symmetry arguments we have demonstrated microscopically that the exciton eigenstates in TMD monolayers are the L- and T-polarized states with the microscopic dipole moment of the exciton (or microscopic current) oriented parallel and perpendicular to its in-plane wavevector. Equations~\eqref{nocap:renorm} and \eqref{cap:renorm} are valid for arbitrary values of the exciton in-plane wavevector $K$, including both the states inside and outside of the light cone. In what follows, however, we will mostly consider the states outside of the light cone, region IV in Fig.~\ref{ris:light_cone} where the induced field decays with the distance from the monolayer and $\varkappa_{i,z} \in \mathbb{R}$. Thus, the parameters $s_i$ in Eq.~\eqref{not:1} and $\xi_i$ in Eq.~\eqref{not:2} are real. As a result, $\omega_{L,T}$ are real. In this case, as expected, the coupling with the induced electromagnetic field, i.e., the long-range exchange interaction between the electron and hole, produces the splitting of the L- and T-exciton energies. Note that for the states inside the light cone the eigenfrequencies $\omega_L$ and $\omega_T$ contain imaginary parts as well being responsible for the radiative damping of excitons, see Refs.~\cite{glazov2014exciton,PhysRevLett.123.067401} for details.

Figure~\ref{ris:lt_splitting} shows the results for the exciton LT-splitting,
\begin{equation}
\label{LT:def}
\Delta E_{LT} = \hbar\omega_L - \hbar\omega_T,
\end{equation}
calculated as function of the exciton wavevector $K$ for different thicknesses of the hBN layers. Panels (a) and (b) show the results for the absent cap layer ($d_0=0$), while panels (c) and (d) shows the results for the structure with the cap layer. Solid lines are calculated after Eqs.~\eqref{nocap:renorm} and \eqref{cap:renorm}, while dotted lines are the analytical asymptotics, Eqs.~\eqref{eq:assympt_lt} and \eqref{eq:assympt_lt:cap}, see below.

Let us first analyze the LT-splitting as a function of the exciton wavevector $K$. At small wavevectors $K \lesssim \omega_0/c$ the $\Delta E_{LT}$ is a strongly non-linear function of $K$ and its real part vanishes for the states within the light cone.\footnote{In the region II there are both real and imaginary parts of $\omega_{L,T}$ due to the leaky waveguide-like modes in the structure.} For sufficiently large exciton wave vectors, $K \gg \omega_0/c$, the parameter $\zeta \sim K_{\parallel}^{-1} \rightarrow 0$. It follows then from Eqs.~(\ref{eq:s-mod}) and \eqref{cap:T} that, $\omega_{T} - \omega_0 \propto \zeta$. Thus, for large wave vectors the energy of the transversal exciton is almost not renormalized. Conversely, one can see from the formulas~\eqref{eq:p-mod} and \eqref{cap:L} that $\omega_{L} - \omega_0 \sim \zeta^{-1} \sim K_{\parallel}$. Therefore, the longitudinal exciton energy renormalization and the LT-splitting of the radiative doublet for large enough exciton wavevectors are equal and linear in $K$.

The asymptotic behavior of the exciton LT-splitting at $K \gg \omega_0/c$ can be recast as
\begin{equation}
    \label{eq:assympt_lt}
        \Delta E_{LT} = \frac{\hbar\Gamma_0}{\varepsilon_{\mathrm{eff}}(K,d_0,d_1)}\frac{cK}{\omega_0},
\end{equation}
with the effective dielectric constant $\varepsilon_{\mathrm{eff}}(K,d_0,d_1)$ being a function of the exciton wavevector and the structure geometry. In the structures with negligible cap layer thickness, $Kd_0 \ll 1$
\begin{equation}
    \label{eq:assympt_lt:nocap}
\varepsilon_{\mathrm{eff}}(K,0,d_1) = \frac{1}{2}\times
\begin{cases}
	{1 + \varepsilon_{\mathrm{SiO_2}}},\quad K d_1 \ll 1,\\
	{1 + \varepsilon_{\mathrm{hBN}}}, \quad K d_1 \gg 1 .
\end{cases}
\end{equation}
Physical sense of this expression is as follows. If there is no substrate hBN layer at all or hBN layer thickness is negligible, i.e, $K d_1 \ll 1$, the field decays mainly into the vacuum and SiO$_2$ substrate. As a result, an effective permittivity of such structure is the average of the permittivities of vacuum and SiO$_2$. When the hBN layer thickness is sufficiently large, $Kd_1 \gg 1$, the electric field induced by the exciton decays into the hBN layer and there is almost no field in SiO$_2$. So effective permittivity contains that of hBN instead SiO$_2$.

This behavior is illustrated in Fig.~\ref{ris:lt_splitting}(a,b). The curve corresponding to the intermediate thickness of the hBN substrate layer, $d_{1} = 9.6$ \AA~(i.e., 3 monoatomic layers of hBN)\footnote{The thickness of monoatomic hexaginal boron nitride equals to $3.2$~\AA.} for small wave vectors is close to zero thickness asymptotics, and for large ones it goes over to the thick hBN layer asymptotics. The behavior of the $\Delta E_{LT}$ for sufficiently large $K$ is very well described by the linear law~\eqref{eq:assympt_lt} with the effective dielectric constant $\varepsilon_{\mathrm{eff}}$ given by Eq.~\eqref{eq:assympt_lt:nocap}. 

Similar results take place in the structures with the cap layer. Assuming that its thickness is sufficiently large, $Kd_0 \gg 1$, we have Eq.~\eqref{eq:assympt_lt} with the effective permittivity in the form
\begin{equation}
    \label{eq:assympt_lt:cap}
\varepsilon_{\mathrm{eff}}(K,\infty,d_1) = \frac{1}{2}\times
\begin{cases}
	{\varepsilon_{\rm hBN} + \varepsilon_{\mathrm{SiO_2}}},\quad K d_1 \ll 1,\\
	{2\varepsilon_{\mathrm{hBN}}}, \quad K d_1 \gg 1.
\end{cases}
\end{equation}
This expression is analogous to Eq.~(\ref{eq:assympt_lt}) except that instead of the vacuum permittivity, which is equal to $1$, the hBN permittivity enters $\varepsilon_{\rm{eff}}$ in Eq.~\eqref{eq:assympt_lt:cap}. This is because for large wavevectors the exciton-induced field is mostly concentrated in the cap hBN layer and does not reach vacuum. Correspoding behavior is illustrated in Fig.~\ref{ris:lt_splitting} panels (c) and (d).

In contrast to the case of the environment effect on the exciton radiative decay rate, where the decay rate shows pronounced oscillations as a function of the hBN layer thickness~\cite{PhysRevLett.123.067401}, here the hBN layer thickness enters $\omega_{L,T}$ through the damped exponential function. The difference is because the LT-splitting of the exciton takes place for the states outside the light cone, where the exciton-induced field decays exponentially to both sides of the monolayer. It is seen from Eqs.~\eqref{nocap:renorm}, \eqref{cap:renorm} and asymptotic expressions~\eqref{eq:assympt_lt}, \eqref{eq:assympt_lt:nocap}, and \eqref{eq:assympt_lt:cap} that increasing the hBN thickness results in a reduction of the LT splitting.

 One can say that the long-range exchange interaction is screened by the presence of the hBN layers. Calculations presented in Fig.~\ref{ris:lt_splitting} confirm this result.

\section{Control of the exciton spin/valley dynamics}\label{sec:dynamics}

In this section we present the model description of the exciton valley dynamics. We present and solve the kinetic equation for the exciton density matrix and analyze the impact of the environment in the van der Waals heterostructure on the valley depolarization.

\subsection{Kinetic equation and its solution}

We describe valley dynamics of excitons in monolayer semiconductors within the pseudospin density matrix approach~\cite{ivchenko05a,glazov2014exciton,PSSB:PSSB201552211,PhysRevB.93.045414}. We introduce the $2\times2$ density matrix
\begin{equation}
\label{dens}
\varrho_{\bm K} = n_{\bm K} + \bm\sigma \cdot \bm s_{\bm K},
\end{equation}
where $n_{\bm K}$ is the average occupancy of the orbital state $\bm K$, i.e., $n_{\bm K}$ is the exciton distribution function, and $\bm s_{\bm K}$ is the pseudospin distribution function, with $s_{\bm K,z}$ component describing the valley polarization or circular polarization of excitons, while the in-plane components $s_{\bm K,x}, s_{\bm K,y}$ describe the valley coherence or exciton alignment/linear polarization. In Eq.~\eqref{dens}, $\bm \sigma = (\sigma_x,\sigma_y,\sigma_z)$ is the vector composed of the $2\times 2$ Pauli matrices; the unit matrix in this notation is omitted.

In the basis of circularly polarized components, the Hamiltonian of the exciton LT-splitting takes the form
\begin{multline}
\label{H:LT}
\mathcal H(\bm K) = \frac{\Delta E_{LT}}{2} \left[\sigma_x \cos{(2\varphi_{\bm K})} + \sigma_y \sin{(2\varphi_{\bm K})} \right] \\= \frac{\hbar}{2} (\bm \Omega_{\bm K}\bm \sigma),
\end{multline}
where the vector 
\begin{equation}
\label{Omega:LT}
\bm \Omega_{\bm K} = (\Delta E_{LT}/\hbar) [\cos{(2\varphi_{\bm K})},\sin{(2\varphi_{\bm K})},0],
\end{equation} 
plays a role of the exciton pseudospin precession frequency in the effective field caused by the LT-splitting.

Within the relaxation time approximation the kinetic equation for the exciton pseudospin distribution takes the form~\cite{PSSB:PSSB201552211} [cf. Ref.~\cite{dyakonov72}]
\begin{equation}
    \label{eq:kinetic}
    \pdv{\vect{s}_{\vect{K}}}{t} + \vect{s}_{\vect{K}}\times\vect{\Omega}_{\vect{K}} + \frac{\vect{s}_{\vect{K}} - \bar{\vect{s}}_{K}}{\tau} =\bm g_{\bm K}.
\end{equation}
Here $\tau$ is the exciton relaxation time, $\bar{\vect{s}}_{{K}}=(2\pi)^{-1}\int_0^{2\pi} \bm s_{\bm K} \dd\varphi_{\bm K}$ is the angular average of the exciton pseudospin, and $\bm g_{\bm K}$ is the pseudospin generation rate.

In what follows we consider the simplest and experimentally relevant situation where the valley-polarized excitons are created by a short circularly polarized light pulse. We perform further calculations in the approximation of the fast exciton energy relaxation: we suppose that after an ensemble of excitons is excited by optical pulse, the Boltzmann energy distribution sets in a short time by valley conserving processes. Thus we employ the following initial condition for Eq.~\eqref{eq:kinetic} and set $\bm g_{\bm K}=0$:
 \begin{equation}
    \label{eq:spin_dist}
    s_{z,K}(t=0) = s_0\frac{2\pi\hbar^2}{m\mathcal S}\frac{\exp(-\epsilon/T)}{T}.
 \end{equation}
Here $\epsilon = \hbar^2 K^2/2m$ is the exciton kinetic energy, $T$ is the temperature measured in energy units ($k_B \equiv 1$), $\mathcal S$ is the normalization area, and $s_0=\sum_{\bm K} s_{z,K}(t=0)$ is the average spin at $t = 0$. Since our aim is to study the effect of the dielectric environment on the exciton valley dynamics we abstain from the description and analysis of the exciton formation processes and details of its energy relaxation, cf. Refs.~\cite{Selig_2018,PhysRevResearch.1.022007,PhysRevB.101.115307,2020arXiv200705369P}. We also stress that the condition $T\tau/\hbar \gg 1$ is fulfilled, otherwise the corrections to the kinetic equation related, e.g, to the weak localization effects should be taken into account~\cite{PhysRevLett.124.166802}.

Making use of the explicit form of $\bm \Omega_{\bm K}$ one arrives at the following equation for the $\bar s_{z,K}$~\cite{glazov2007}:
\begin{equation}
\label{sz:kin}
\left(\frac{\partial}{\partial t}  + \frac{1}{\tau}\right)\frac{\partial}{\partial t} \bar s_{z,K} + \Omega_K^2 \bar s_{z,K} =0,
\end{equation}
where we took into account that $s_{z,\bm K} = \bar s_{z,K}$. In agreement with Refs.~\cite{PhysRevLett.81.2586,gridnev01} we obtain the expression for valley polarization dynamics:
 \begin{equation}
    \label{eq:spin_pol_unav}
    \bar s_{z, K} = e^{-\frac{t}{2\tau}}\left(\frac{\sinh\frac{qt}{2\tau}}{q} + \cosh\frac{qt}{2\tau}\right) s_{z,K}(t=0),
 \end{equation}
 where $q \equiv q(\epsilon) = \sqrt{1 - \left(2\Delta E_{LT}\tau/\hbar\right)^2}$. Ultimately, we arrive at the following
 expression for valley polarization dynamics of excitions:
\begin{multline}
    \label{eq:spin_pol}
	S_z(t) =\sum_{\bm K} \bar s_{z,K}\\
	= s_0e^{-\frac{t}{2\tau}}\int\limits_0^{\infty}\frac{e^{-\frac{\epsilon}{T}}}{T}\left(\frac{\sinh\frac{q(\epsilon)t}{2\tau}}{q(\epsilon)} + \cosh\frac{q(\epsilon)t}{2\tau}\right)\dd\epsilon.
\end{multline}
Strictly speaking, the integral over energy in Eq.~\eqref{eq:spin_pol} should be cut-off at small energies $\epsilon^* \sim \hbar^2 \omega_0^2/(2mc^2)$, i.e., for the states within the light cone. Estimates show that this cut-off is unimportant at reasonable temperatures $T\gtrsim 1$~K.

%\newpage

\subsection{Exciton spin/valley dynamics in limiting cases}

Before turning to the numerical results, let us deduce analytical asymptotics of the integral~\eqref{eq:spin_pol} for important limiting cases. Characteristic -- average -- energy of the exciton ensemble is the temperature $T$, which determines a typical value of the thermal wavevector $K_T = \sqrt{2mT}/\hbar$ and, accordingly, the typical pseudospin precession frequency $\Omega_T \equiv \Omega_{K_T}$. We consider the behaviour of the integral~\eqref{eq:spin_pol} and spin dynamics in the two important cases, where the pseudospin precession frequency is either much smaller than the scattering rate $\tau^{-1}$
\begin{subequations}
\label{crit:Omega}
\begin{equation}
\label{frequent}
\Omega_T \tau \ll 1,
\end{equation}
or much larger than $\tau^{-1}$:
\begin{equation}
\label{rare}
\Omega_T \tau \gg 1.
\end{equation}
\end{subequations}
In both cases simple analytical expressions describing the spin dynamics are derived.

\begin{figure}[tb]
\centering
\includegraphics[width=\linewidth]{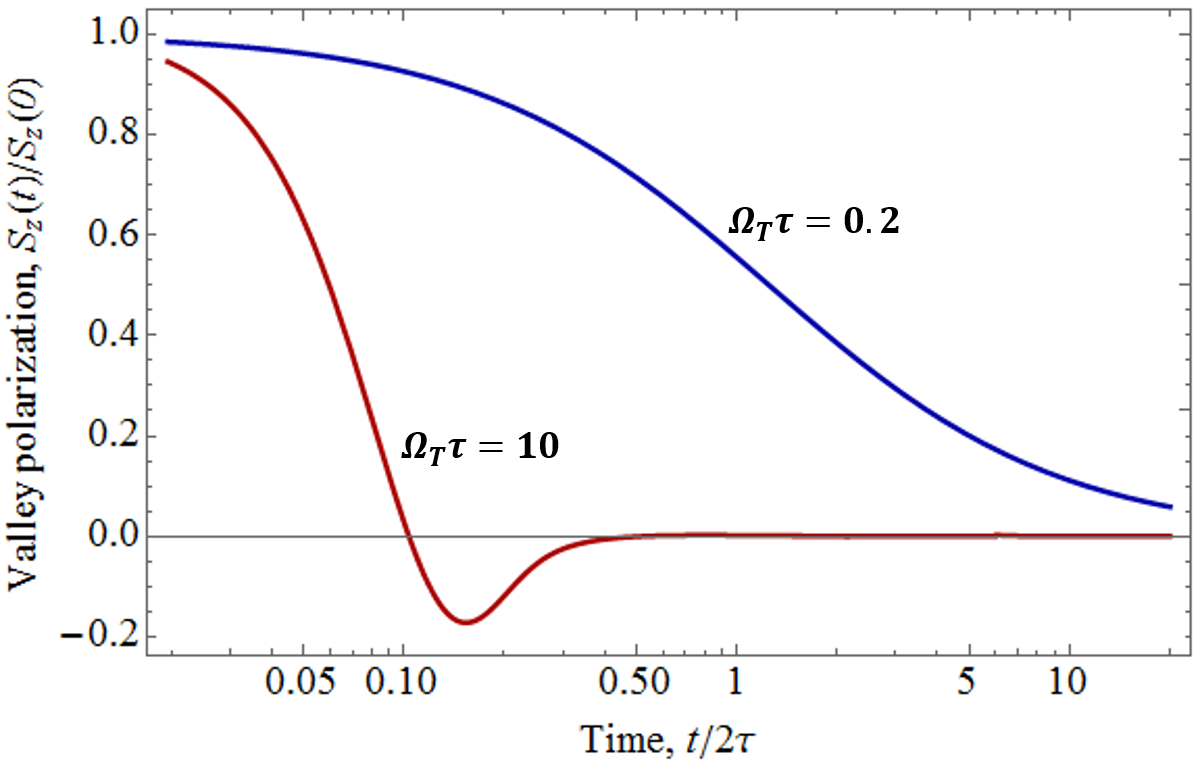}
\caption{\textbf{Exciton spin/valley dynamics in limiting cases.} Blue curve illustrates characteristic valley polarization dynamics in collision-dominated regime ($\Omega_T\tau \ll 1$), red one corresponds to the rare scattering regime ($\Omega_T \tau \gg 1$).}
\label{ris:spin_asymp}
\end{figure}

\begin{figure*}[t]
\includegraphics[width=\linewidth]{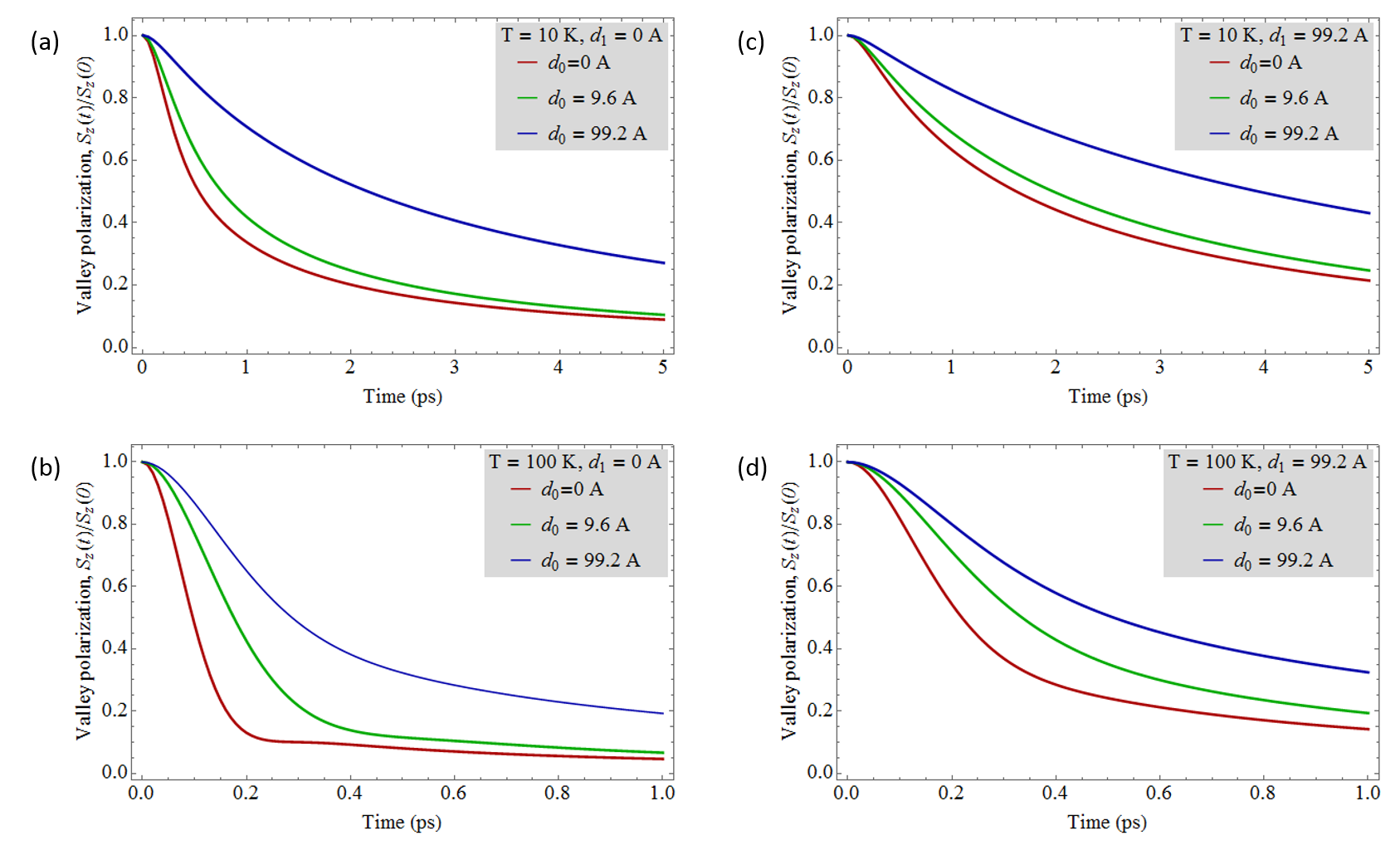}
\caption{\textbf{Valley polarization dynamics for different structure parameters.} 
Panels (a) and (b) correspond to the absent substrate hBN layer ($d_1=0$) and two temperatures $T=10$~K and $100$~K, respectively. Panels (c) and (d) correspond to the structure with sufficiently thick substrate hBN layer ($d_1=99.2$~\AA) and two temperatures $T=10$~K and $100$~K. Different curves show the valley polarization dynamics for different top hBN layer thicknesses. Parameters of the calculations are the same as in Fig.~\ref{ris:lt_splitting} and the scattering time $\tau = 0.1$ ps is assumed to be temperature and energy independent.}
\label{ris:experimentalcorrelationsignals}
\end{figure*}

In the first situation where the scattering acts are frequent, Eq.~\eqref{frequent}, we use the asymptotics $q(\epsilon) \approx 1 - 2\left(\Delta E_{LT}\tau/\hbar\right)^2$, and obtain:
\begin{equation}
\label{dyn:frequent}
    S_z(t) = s_0\int\limits_0^{\infty}\frac{e^{-\frac{\epsilon}{T}}}{T}\exp\left[-(\Delta E_{LT}/\hbar)^2\tau t\right]\dd\epsilon.
\end{equation}
The energy dependence of the subintegral expression results from both the Boltzmann exponent $\exp(-\epsilon/T)$ and the wavevector dependence on the $\Delta E_{LT}$. The latter can be written, in accordance with Eq.~\eqref{eq:assympt_lt}, as
\begin{equation}
\label{DELT:mod}
\Delta E_{LT} = \hbar \beta \sqrt{\varepsilon}, \quad \beta = \sqrt{\frac{2m c^2}{\omega_0^2}} \frac{\Gamma_0}{\varepsilon_{\rm eff}(K_T,d_0,d_1)}.
\end{equation}
In derivation of Eq.~\eqref{DELT:mod} we disregarded $K$-dependence of the effective permittivity assuming that relevant wavevectors are sufficiently large. The resulting integral is readily evaluated as

\begin{equation}
    \label{eq:asympt_spin_ll}
    S_z(t) = \frac{s_0}{1 + T\beta^2\tau t}.
\end{equation}
The exciton valley depolarization rate is given by 
\begin{equation}
\label{valley:DP}
\frac{1}{\tau_v} \equiv \beta^2 T \tau \sim \Omega_T^2\tau,
\end{equation} 
in accordance with the general result in the collision-dominated regime~\cite{dyakonov72,glazov2014exciton,PhysRevB.90.161302}. Interestingly, the decay is slow with $S_z(t) \propto t^{-1}$ at $t \gg \tau_v$. This $1/t$ `tail' is a result of neglected energy relaxation processes. If the exciton energy relaxation time $\tau_\epsilon$ is sufficiently short as compared with the valley depolarization time, $\tau_\epsilon \ll \tau_v$, but simultaneously sufficiently long compared to the momentum relaxation time, $\tau_\epsilon \gg \tau$, then the exciton ensemble is characterized by a single relaxation rate $\tau_v^{-1}$ in Eq.~\eqref{valley:DP} and, instead of Eqs.~\eqref{dyn:frequent} and \eqref{eq:asympt_spin_ll} we obtain 
\begin{multline}
\label{dyn:frequent:alt}
    S_z(t) = s_0\exp\left[-\int\limits_0^{\infty}\frac{e^{-\frac{\epsilon}{T}}}{T}(\Delta E_{LT}/\hbar)^2\tau t\dd\epsilon\right]\\ 
    = s_0 \exp(-t/\tau_v).
\end{multline}
In this situation standard exponential decay law of the valley polarization is recovered.

Now we turn to the limit of rare scattering events~\cite{gridnev01,brand02}, Eq.~\eqref{rare} we have $q(\epsilon) \approx 2i\Delta E_{LT}\tau/\hbar$, and it follows from Eqs.~\eqref{eq:spin_pol} and \eqref{DELT:mod} that [cf. Ref.~\cite{PSSB:PSSB201552211}]
\begin{multline}
    \label{eq:asympt_spin_gg}
    S_z(t) = s_0e^{-\frac{t}{2\tau}}\int\limits_0^{\infty}\frac{e^{-\frac{\epsilon}{T}}}{T}\cos\left(\beta\sqrt{\varepsilon}t\right)\dd\epsilon 
    \\
    = s_0e^{-\frac{t}{2\tau}}\left[1 - \sqrt{T}\beta t\,\textrm{F}\left(\frac{\sqrt{T}\beta t}{2}\right)\right],
\end{multline}
where $\textrm{F}(x) = \exp(-x^2)\int_0^x\exp(t^2)\dd t$ is the Dawson function. In this regime the valley polarization decays mainly due to the spread of the pseudospin precession frequencies with the characteristic rate $\beta \sqrt{T} \sim \Omega_T$. The scattering breaks phase of the pseudospin precession and results in the additional exponential decay with the rate $1/(2\tau)$~\cite{gridnev01}.

From asymptotics~\eqref{eq:asympt_spin_ll} and~\eqref{eq:asympt_spin_gg} one can see, that in the case $\Omega_{T}\tau \ll 1$ we expect slow monotonous relaxation of the exciton valley polarization as shown in Fig.~\ref{ris:spin_asymp}. If, by contrast, $\Omega_{T}\tau \gg 1$, we expect fast decoherence with a characteristic minima, see Fig.~\ref{ris:spin_asymp}.

\subsection{Numerical results and discussion}

Figure~\ref{ris:experimentalcorrelationsignals} shows the valley polarization dynamics calculated numerically after Eq.~\eqref{eq:spin_pol} for various parameters of the structure and two characteristic temperatures $T=10$~K and $100$~K. Panels (a) and (b) show the dynamics for the structure without a substrate hBN layer, while panels (c) and (d) demonstrate the dynamics in the structures with the substrate layer. Overall behavior of the valley polarization $S_z(t)$ is intermediate between the asymptotics shown in Fig.~\ref{ris:spin_asymp}. Figure~\ref{ris:experimentalcorrelationsignals} demonstrates clearly that the exciton valley relaxation time can be controlled by the dielectric environment engineering.

For a fixed hBN layers thicknesses the valley depolarization rate increases with increase of the temperature. This is because at a higher temperature the characteristic pseudospin precession frequency $\Omega_T$ increases. It is in agreement with experimental data~\cite{PhysRevB.90.161302}.

An increase of the hBN layers thicknesses results, as discussed in Sec.~\ref{sec:fine}, in the effective screening of the exchange interaction and, correspondingly, in suppression of the exciton LT-splitting. As a result, at a fixed thickness of the substrate hBN layer $d_1$, an increase in the cap layer thickness $d_0$ slows-down the valley depolarization, compare red, green and blue curves in the panels. Similarly, an increase in $d_1$ at a fixed $d_0$ slows down depolarization as well, compare Fig.~\ref{ris:experimentalcorrelationsignals} (a) with (c) and (b) with (d).

Calculations show that the spin dynamics is fastest for the structure without hBN, $d_0=d_1=0$. At $T=100$~K the product $\Omega_T\tau$ exceeds unity and slightly non-monotonic behavior of the red curve in Fig.~\ref{ris:experimentalcorrelationsignals}(b) is seen. 
Overall, the modulation of the valley depolariza- tion time for different system parameters is significant,  compare red and blue lines in Fig.~\ref{ris:experimentalcorrelationsignals}(a).
%Overall, a modulation of the valley depolarization time for different system parameters reaches \addMisha{several} times, compare red and blue lines in Fig.~\ref{ris:experimentalcorrelationsignals}(a).

\begin{figure}[htb]
\centering
\includegraphics[width=\linewidth]{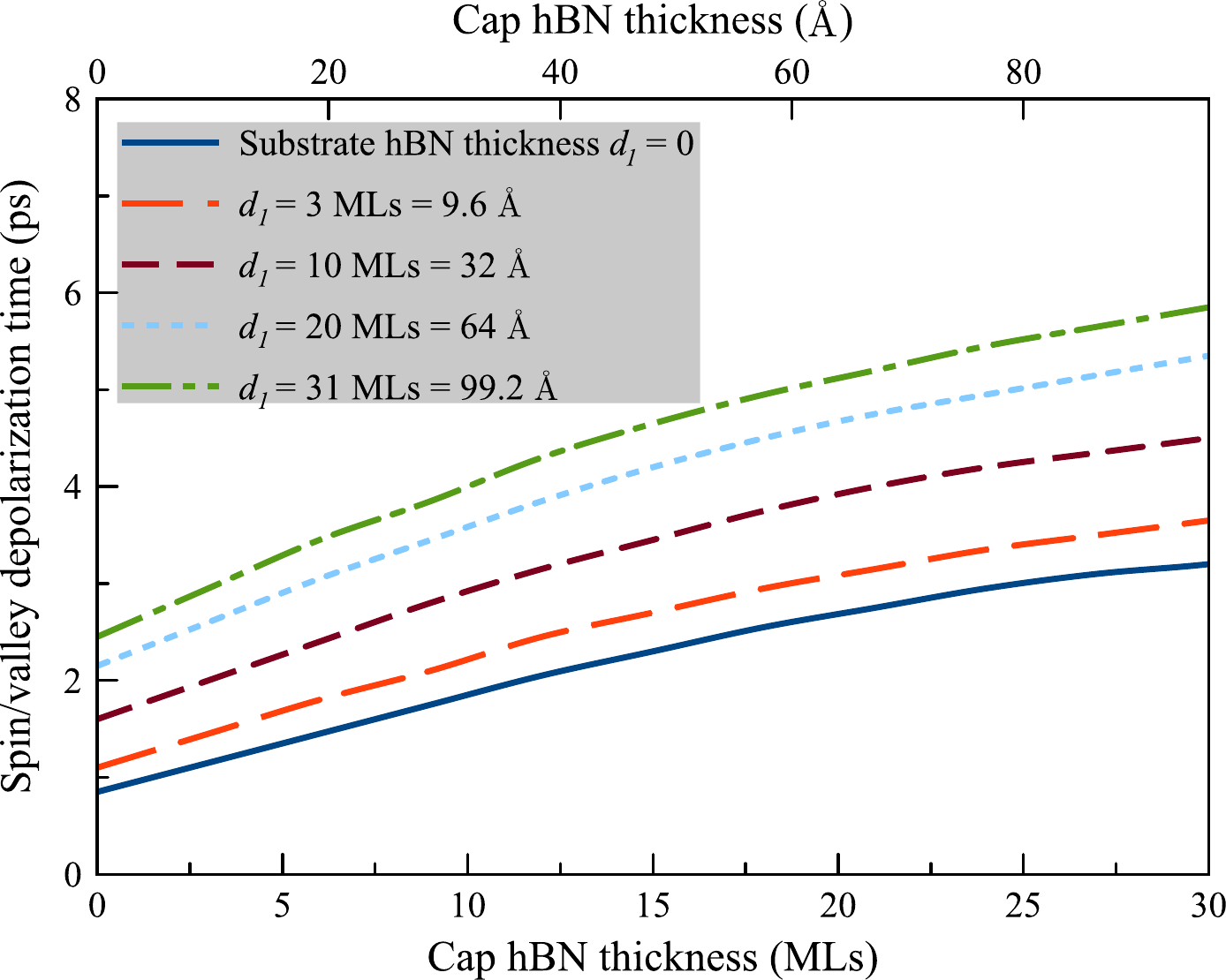}
\caption{\textbf{Controlling the exciton spin/valley depolarization.} Exciton spin/valley polarization lifetime $\tau_v$ as a function of the top hBN layer thickness $d_0$ for the structure shown in Fig.~\ref{ris:stuctures}(b) at calculated for different values of the substrate hBN thickness $d_1$ (different curves). Temperature $T=10$~K, scattering time $\tau=0.1$~ps. The depolarization time $\tau_v$ is defined as $S_z(\tau_v)/S_z(0)=1/e$. }
\label{ris:control}
\end{figure}

The predictions for the control of the exciton spin/valley polarization lifetime are summarized in Fig.~\ref{ris:control} where the dependence of the $\tau_v$ on the cap hBN layer thickness is presented for the structure shown in Fig.~\ref{ris:stuctures}(b) for different substrate hBN thicknesses $d_1$. We determine the spin/valley depolarization time $\tau_v$ from the condition $S_z(\tau_v)/S_z(0)=1/e$, i.e., it corresponds to the decay by $e \approx 2.718$. One can see that for a fixed $d_1$ the depolarization time increases with increasing $d_0$ and, similarly, for a fixed $d_0$ the depolarization time increases with increasing $d_1$. This is because of the effective screening of the electron-hole long-range exchange interaction. The significant modulation of $\tau_v$ is seen. Note that significant variation of $\tau_v$ is observed for very small variations ($\sim$~nm) of the hBN thickness. Comparing the structures without encapsulation $d_1=d_0=0$ and structures with sufficiently thick encapsulation, $30$~MLs of hBN for both the cap and substrate layers one can see that the variation of $\tau_v$ by a factor $\gtrsim 5$ is possible.

\section{Conclusion}\label{sec:concl}

We have studied the effect of the dielectric environment of the atomically thin semiconductor on the exciton fine structure and its valley depolarization in van der Waals heterostructures based on transition metal dichalcogenide monolayers encapsulated into hexagonal boron nitride. The microscopic theory of the exciton fine structure has been developed within the electrodynamical approach where the long-range exchange interaction naturally appears as a result of the exciton coupling with the induced electromagnetic field. The valley dynamics has been studied within the kinetic equation approach for the pseudospin density matrix.

We have demonstrated that the encapsulation of the monolayer into hBN effectively screens the long-range exchange interaction and results in a slow-down of the valley depolarization. While the radiative decay of excitons in monolayer semiconductors and the electron-hole long-range exchange interaction have the same physical origin, related to the self-consistent interaction of the exciton with its electromagnetic field, their dependence on the boron nitride layer thickness is different. In the radiative recombination process, the excitonic states within the light cone are involved. Those states induce propagating electromagnetic field which oscillates in space. As a result, the radiative decay rate shows oscillations as a function of the hBN thickness~\cite{PhysRevLett.123.067401}. In the studied case of the long-range exchange interaction, the excitons are outside of the light cone and they induce decaying in space electromagnetic field. It gives rise to a monotinic dependence of the longitudinal-transverse splitting of excitonic states as a function of the hBN thickness. 

Our calculations demonstrate a significant, up to five-fold, variation of the valley depolarization in hBN-based van der Waals heterostructures. Our results open up the possibilities to control the exciton valley dynamics by appropriately tailoring the electrodynamical environment of the monolayer.

\acknowledgments

We thank M.A. Semina for valuable discussions. This work was partially supported by the Russian Science Foundation, project No. 19-12-00051,  ANR  2D-vdW-Spin,  MagicValley, and Vallex, as well as the French-Russian IRP PHYNICS. A.I.P. was supported in part by the Foundation for the Advancement of Theoretical Physics and Mathematics ``BASIS''.


\begin{thebibliography}{55}%
\makeatletter
\providecommand \@ifxundefined [1]{%
 \@ifx{#1\undefined}
}%
\providecommand \@ifnum [1]{%
 \ifnum #1\expandafter \@firstoftwo
 \else \expandafter \@secondoftwo
 \fi
}%
\providecommand \@ifx [1]{%
 \ifx #1\expandafter \@firstoftwo
 \else \expandafter \@secondoftwo
 \fi
}%
\providecommand \natexlab [1]{#1}%
\providecommand \enquote  [1]{``#1''}%
\providecommand \bibnamefont  [1]{#1}%
\providecommand \bibfnamefont [1]{#1}%
\providecommand \citenamefont [1]{#1}%
\providecommand \href@noop [0]{\@secondoftwo}%
\providecommand \href [0]{\begingroup \@sanitize@url \@href}%
\providecommand \@href[1]{\@@startlink{#1}\@@href}%
\providecommand \@@href[1]{\endgroup#1\@@endlink}%
\providecommand \@sanitize@url [0]{\catcode `\\12\catcode `\$12\catcode
  `\&12\catcode `\#12\catcode `\^12\catcode `\_12\catcode `\%12\relax}%
\providecommand \@@startlink[1]{}%
\providecommand \@@endlink[0]{}%
\providecommand \url  [0]{\begingroup\@sanitize@url \@url }%
\providecommand \@url [1]{\endgroup\@href {#1}{\urlprefix }}%
\providecommand \urlprefix  [0]{URL }%
\providecommand \Eprint [0]{\href }%
\providecommand \doibase [0]{http://dx.doi.org/}%
\providecommand \selectlanguage [0]{\@gobble}%
\providecommand \bibinfo  [0]{\@secondoftwo}%
\providecommand \bibfield  [0]{\@secondoftwo}%
\providecommand \translation [1]{[#1]}%
\providecommand \BibitemOpen [0]{}%
\providecommand \bibitemStop [0]{}%
\providecommand \bibitemNoStop [0]{.\EOS\space}%
\providecommand \EOS [0]{\spacefactor3000\relax}%
\providecommand \BibitemShut  [1]{\csname bibitem#1\endcsname}%
\let\auto@bib@innerbib\@empty
%</preamble>
\bibitem [{\citenamefont {Geim}\ and\ \citenamefont
  {Grigorieva}(2013)}]{Geim:2013aa}%
  \BibitemOpen
  \bibfield  {author} {\bibinfo {author} {\bibfnamefont {A.~K.}\ \bibnamefont
  {Geim}}\ and\ \bibinfo {author} {\bibfnamefont {I.~V.}\ \bibnamefont
  {Grigorieva}},\ }\bibfield  {title} {\enquote {\bibinfo {title} {Van der
  $\mbox{W}$aals heterostructures},}\ }\href
  {http://dx.doi.org/10.1038/nature12385} {\bibfield  {journal} {\bibinfo
  {journal} {Nature}\ }\textbf {\bibinfo {volume} {499}},\ \bibinfo {pages}
  {419--425} (\bibinfo {year} {2013})}\BibitemShut {NoStop}%
\bibitem [{\citenamefont {Butler}\ \emph {et~al.}(2013)\citenamefont {Butler},
  \citenamefont {Hollen}, \citenamefont {Cao}, \citenamefont {Cui},
  \citenamefont {Gupta}, \citenamefont {Guti\'{e}rrez}, \citenamefont {Heinz},
  \citenamefont {Hong}, \citenamefont {Huang}, \citenamefont {Ismach},
  \citenamefont {Johnston-Halperin}, \citenamefont {Kuno}, \citenamefont
  {Plashnitsa}, \citenamefont {Robinson}, \citenamefont {Ruoff}, \citenamefont
  {Salahuddin}, \citenamefont {Shan}, \citenamefont {Shi}, \citenamefont
  {Spencer}, \citenamefont {Terrones}, \citenamefont {Windl},\ and\
  \citenamefont {Goldberger}}]{Butler:2013a}%
  \BibitemOpen
  \bibfield  {author} {\bibinfo {author} {\bibfnamefont {Sheneve~Z.}\
  \bibnamefont {Butler}}, \bibinfo {author} {\bibfnamefont {Shawna~M.}\
  \bibnamefont {Hollen}}, \bibinfo {author} {\bibfnamefont {Linyou}\
  \bibnamefont {Cao}}, \bibinfo {author} {\bibfnamefont {Yi}~\bibnamefont
  {Cui}}, \bibinfo {author} {\bibfnamefont {Jay~A.}\ \bibnamefont {Gupta}},
  \bibinfo {author} {\bibfnamefont {Humberto~R.}\ \bibnamefont
  {Guti\'{e}rrez}}, \bibinfo {author} {\bibfnamefont {Tony~F.}\ \bibnamefont
  {Heinz}}, \bibinfo {author} {\bibfnamefont {Seung~Sae}\ \bibnamefont {Hong}},
  \bibinfo {author} {\bibfnamefont {Jiaxing}\ \bibnamefont {Huang}}, \bibinfo
  {author} {\bibfnamefont {Ariel~F.}\ \bibnamefont {Ismach}}, \bibinfo {author}
  {\bibfnamefont {Ezekiel}\ \bibnamefont {Johnston-Halperin}}, \bibinfo
  {author} {\bibfnamefont {Masaru}\ \bibnamefont {Kuno}}, \bibinfo {author}
  {\bibfnamefont {Vladimir~V.}\ \bibnamefont {Plashnitsa}}, \bibinfo {author}
  {\bibfnamefont {Richard~D.}\ \bibnamefont {Robinson}}, \bibinfo {author}
  {\bibfnamefont {Rodney~S.}\ \bibnamefont {Ruoff}}, \bibinfo {author}
  {\bibfnamefont {Sayeef}\ \bibnamefont {Salahuddin}}, \bibinfo {author}
  {\bibfnamefont {Jie}\ \bibnamefont {Shan}}, \bibinfo {author} {\bibfnamefont
  {Li}~\bibnamefont {Shi}}, \bibinfo {author} {\bibfnamefont {Michael~G.}\
  \bibnamefont {Spencer}}, \bibinfo {author} {\bibfnamefont {Mauricio}\
  \bibnamefont {Terrones}}, \bibinfo {author} {\bibfnamefont {Wolfgang}\
  \bibnamefont {Windl}}, \ and\ \bibinfo {author} {\bibfnamefont {Joshua~E.}\
  \bibnamefont {Goldberger}},\ }\bibfield  {title} {\enquote {\bibinfo {title}
  {Progress, challenges, and opportunities in two-dimensional materials beyond
  graphene},}\ }\href {\doibase 10.1021/nn400280c} {\bibfield  {journal}
  {\bibinfo  {journal} {ACS Nano}\ }\textbf {\bibinfo {volume} {7}},\ \bibinfo
  {pages} {2898--2926} (\bibinfo {year} {2013})}\BibitemShut {NoStop}%
\bibitem [{\citenamefont {Splendiani}\ \emph {et~al.}(2010)\citenamefont
  {Splendiani}, \citenamefont {Sun}, \citenamefont {Zhang}, \citenamefont {Li},
  \citenamefont {Kim}, \citenamefont {Chim}, \citenamefont {Galli},\ and\
  \citenamefont {Wang}}]{Splendiani:2010a}%
  \BibitemOpen
  \bibfield  {author} {\bibinfo {author} {\bibfnamefont {Andrea}\ \bibnamefont
  {Splendiani}}, \bibinfo {author} {\bibfnamefont {Liang}\ \bibnamefont {Sun}},
  \bibinfo {author} {\bibfnamefont {Yuanbo}\ \bibnamefont {Zhang}}, \bibinfo
  {author} {\bibfnamefont {Tianshu}\ \bibnamefont {Li}}, \bibinfo {author}
  {\bibfnamefont {Jonghwan}\ \bibnamefont {Kim}}, \bibinfo {author}
  {\bibfnamefont {Chi-Yung}\ \bibnamefont {Chim}}, \bibinfo {author}
  {\bibfnamefont {Giulia}\ \bibnamefont {Galli}}, \ and\ \bibinfo {author}
  {\bibfnamefont {Feng}\ \bibnamefont {Wang}},\ }\bibfield  {title} {\enquote
  {\bibinfo {title} {Emerging photoluminescence in monolayer {MoS}$_2$},}\
  }\href@noop {} {\bibfield  {journal} {\bibinfo  {journal} {Nano Letters}\
  }\textbf {\bibinfo {volume} {10}},\ \bibinfo {pages} {1271} (\bibinfo {year}
  {2010})}\BibitemShut {NoStop}%
\bibitem [{\citenamefont {Mak}\ \emph {et~al.}(2010)\citenamefont {Mak},
  \citenamefont {Lee}, \citenamefont {Hone}, \citenamefont {Shan},\ and\
  \citenamefont {Heinz}}]{Mak:2010bh}%
  \BibitemOpen
  \bibfield  {author} {\bibinfo {author} {\bibfnamefont {Kin~Fai}\ \bibnamefont
  {Mak}}, \bibinfo {author} {\bibfnamefont {Changgu}\ \bibnamefont {Lee}},
  \bibinfo {author} {\bibfnamefont {James}\ \bibnamefont {Hone}}, \bibinfo
  {author} {\bibfnamefont {Jie}\ \bibnamefont {Shan}}, \ and\ \bibinfo {author}
  {\bibfnamefont {Tony~F.}\ \bibnamefont {Heinz}},\ }\bibfield  {title}
  {\enquote {\bibinfo {title} {Atomically thin {MoS}$_{2}$: A new direct-gap
  semiconductor},}\ }\href {\doibase 10.1103/PhysRevLett.105.136805} {\bibfield
   {journal} {\bibinfo  {journal} {Phys. Rev. Lett.}\ }\textbf {\bibinfo
  {volume} {105}},\ \bibinfo {pages} {136805} (\bibinfo {year}
  {2010})}\BibitemShut {NoStop}%
\bibitem [{\citenamefont {He}\ \emph {et~al.}(2014)\citenamefont {He},
  \citenamefont {Kumar}, \citenamefont {Zhao}, \citenamefont {Wang},
  \citenamefont {Mak}, \citenamefont {Zhao},\ and\ \citenamefont
  {Shan}}]{PhysRevLett.113.026803}%
  \BibitemOpen
  \bibfield  {author} {\bibinfo {author} {\bibfnamefont {Keliang}\ \bibnamefont
  {He}}, \bibinfo {author} {\bibfnamefont {Nardeep}\ \bibnamefont {Kumar}},
  \bibinfo {author} {\bibfnamefont {Liang}\ \bibnamefont {Zhao}}, \bibinfo
  {author} {\bibfnamefont {Zefang}\ \bibnamefont {Wang}}, \bibinfo {author}
  {\bibfnamefont {Kin~Fai}\ \bibnamefont {Mak}}, \bibinfo {author}
  {\bibfnamefont {Hui}\ \bibnamefont {Zhao}}, \ and\ \bibinfo {author}
  {\bibfnamefont {Jie}\ \bibnamefont {Shan}},\ }\bibfield  {title} {\enquote
  {\bibinfo {title} {Tightly bound excitons in monolayer $\mbox{WSe}_{2}$},}\
  }\href {\doibase 10.1103/PhysRevLett.113.026803} {\bibfield  {journal}
  {\bibinfo  {journal} {Phys. Rev. Lett.}\ }\textbf {\bibinfo {volume} {113}},\
  \bibinfo {pages} {026803} (\bibinfo {year} {2014})}\BibitemShut {NoStop}%
\bibitem [{\citenamefont {Chernikov}\ \emph {et~al.}(2014)\citenamefont
  {Chernikov}, \citenamefont {Berkelbach}, \citenamefont {Hill}, \citenamefont
  {Rigosi}, \citenamefont {Li}, \citenamefont {Aslan}, \citenamefont
  {Reichman}, \citenamefont {Hybertsen},\ and\ \citenamefont
  {Heinz}}]{Chernikov:2014a}%
  \BibitemOpen
  \bibfield  {author} {\bibinfo {author} {\bibfnamefont {Alexey}\ \bibnamefont
  {Chernikov}}, \bibinfo {author} {\bibfnamefont {Timothy~C.}\ \bibnamefont
  {Berkelbach}}, \bibinfo {author} {\bibfnamefont {Heather~M.}\ \bibnamefont
  {Hill}}, \bibinfo {author} {\bibfnamefont {Albert}\ \bibnamefont {Rigosi}},
  \bibinfo {author} {\bibfnamefont {Yilei}\ \bibnamefont {Li}}, \bibinfo
  {author} {\bibfnamefont {Ozgur~Burak}\ \bibnamefont {Aslan}}, \bibinfo
  {author} {\bibfnamefont {David~R.}\ \bibnamefont {Reichman}}, \bibinfo
  {author} {\bibfnamefont {Mark~S.}\ \bibnamefont {Hybertsen}}, \ and\ \bibinfo
  {author} {\bibfnamefont {Tony~F.}\ \bibnamefont {Heinz}},\ }\bibfield
  {title} {\enquote {\bibinfo {title} {Exciton binding energy and nonhydrogenic
  $\mbox{R}$ydberg series in monolayer $\mbox{WS}_{2}$},}\ }\href {\doibase
  10.1103/PhysRevLett.113.076802} {\bibfield  {journal} {\bibinfo  {journal}
  {Phys. Rev. Lett.}\ }\textbf {\bibinfo {volume} {113}},\ \bibinfo {pages}
  {076802} (\bibinfo {year} {2014})}\BibitemShut {NoStop}%
\bibitem [{\citenamefont {Yu}\ \emph {et~al.}(2015)\citenamefont {Yu},
  \citenamefont {Cui}, \citenamefont {Xu},\ and\ \citenamefont
  {Yao}}]{Yu30122014}%
  \BibitemOpen
  \bibfield  {author} {\bibinfo {author} {\bibfnamefont {Hongyi}\ \bibnamefont
  {Yu}}, \bibinfo {author} {\bibfnamefont {Xiaodong}\ \bibnamefont {Cui}},
  \bibinfo {author} {\bibfnamefont {Xiaodong}\ \bibnamefont {Xu}}, \ and\
  \bibinfo {author} {\bibfnamefont {Wang}\ \bibnamefont {Yao}},\ }\bibfield
  {title} {\enquote {\bibinfo {title} {Valley excitons in two-dimensional
  semiconductors},}\ }\href {\doibase 10.1093/nsr/nwu078} {\bibfield  {journal}
  {\bibinfo  {journal} {National Science Review}\ }\textbf {\bibinfo {volume}
  {2}},\ \bibinfo {pages} {57--70} (\bibinfo {year} {2015})}\BibitemShut
  {NoStop}%
\bibitem [{\citenamefont {Kolobov}\ and\ \citenamefont
  {Tominaga}(2016)}]{Kolobov2016book}%
  \BibitemOpen
  \bibfield  {author} {\bibinfo {author} {\bibfnamefont {Alexander~V.}\
  \bibnamefont {Kolobov}}\ and\ \bibinfo {author} {\bibfnamefont {Junji}\
  \bibnamefont {Tominaga}},\ }\href {\doibase 10.1007/978-3-319-31450-1} {\emph
  {\bibinfo {title} {Two-Dimensional Transition-Metal Dichalcogenides}}}\
  (\bibinfo  {publisher} {Springer International Publishing},\ \bibinfo {year}
  {2016})\BibitemShut {NoStop}%
\bibitem [{\citenamefont {Wang}\ \emph {et~al.}(2018)\citenamefont {Wang},
  \citenamefont {Chernikov}, \citenamefont {Glazov}, \citenamefont {Heinz},
  \citenamefont {Marie}, \citenamefont {Amand},\ and\ \citenamefont
  {Urbaszek}}]{RevModPhys.90.021001}%
  \BibitemOpen
  \bibfield  {author} {\bibinfo {author} {\bibfnamefont {Gang}\ \bibnamefont
  {Wang}}, \bibinfo {author} {\bibfnamefont {Alexey}\ \bibnamefont
  {Chernikov}}, \bibinfo {author} {\bibfnamefont {Mikhail~M.}\ \bibnamefont
  {Glazov}}, \bibinfo {author} {\bibfnamefont {Tony~F.}\ \bibnamefont {Heinz}},
  \bibinfo {author} {\bibfnamefont {Xavier}\ \bibnamefont {Marie}}, \bibinfo
  {author} {\bibfnamefont {Thierry}\ \bibnamefont {Amand}}, \ and\ \bibinfo
  {author} {\bibfnamefont {Bernhard}\ \bibnamefont {Urbaszek}},\ }\bibfield
  {title} {\enquote {\bibinfo {title} {Colloquium: Excitons in atomically thin
  transition metal dichalcogenides},}\ }\href {\doibase
  10.1103/RevModPhys.90.021001} {\bibfield  {journal} {\bibinfo  {journal}
  {Rev. Mod. Phys.}\ }\textbf {\bibinfo {volume} {90}},\ \bibinfo {pages}
  {021001} (\bibinfo {year} {2018})}\BibitemShut {NoStop}%
\bibitem [{\citenamefont {Durnev}\ and\ \citenamefont
  {Glazov}(2018)}]{Durnev_2018}%
  \BibitemOpen
  \bibfield  {author} {\bibinfo {author} {\bibfnamefont {M~V}\ \bibnamefont
  {Durnev}}\ and\ \bibinfo {author} {\bibfnamefont {M~M}\ \bibnamefont
  {Glazov}},\ }\bibfield  {title} {\enquote {\bibinfo {title} {Excitons and
  trions in two-dimensional semiconductors based on transition metal
  dichalcogenides},}\ }\href {\doibase 10.3367/ufne.2017.07.038172} {\bibfield
  {journal} {\bibinfo  {journal} {Physics-Uspekhi}\ }\textbf {\bibinfo {volume}
  {61}},\ \bibinfo {pages} {825--845} (\bibinfo {year} {2018})}\BibitemShut
  {NoStop}%
\bibitem [{\citenamefont {Li}\ \emph {et~al.}(2020)\citenamefont {Li},
  \citenamefont {Wang}, \citenamefont {Miao}, \citenamefont {Lian},\ and\
  \citenamefont {Shi}}]{Li2020}%
  \BibitemOpen
  \bibfield  {author} {\bibinfo {author} {\bibfnamefont {Zhipeng}\ \bibnamefont
  {Li}}, \bibinfo {author} {\bibfnamefont {Tianmeng}\ \bibnamefont {Wang}},
  \bibinfo {author} {\bibfnamefont {Shengnan}\ \bibnamefont {Miao}}, \bibinfo
  {author} {\bibfnamefont {Zhen}\ \bibnamefont {Lian}}, \ and\ \bibinfo
  {author} {\bibfnamefont {Su-Fei}\ \bibnamefont {Shi}},\ }\bibfield  {title}
  {\enquote {\bibinfo {title} {Fine structures of valley-polarized excitonic
  states in monolayer transitional metal dichalcogenides},}\ }\href {\doibase
  10.1515/nanoph-2020-0054} {\bibfield  {journal} {\bibinfo  {journal}
  {Nanophotonics}\ }\textbf {\bibinfo {volume} {9}},\ \bibinfo {pages}
  {1811--1829} (\bibinfo {year} {2020})}\BibitemShut {NoStop}%
\bibitem [{\citenamefont {Raja}\ \emph {et~al.}(2017)\citenamefont {Raja},
  \citenamefont {Chaves}, \citenamefont {Yu}, \citenamefont {Arefe},
  \citenamefont {Hill}, \citenamefont {Rigosi}, \citenamefont {Berkelbach},
  \citenamefont {Nagler}, \citenamefont {Sch{\"u}ller}, \citenamefont {Korn},
  \citenamefont {Nuckolls}, \citenamefont {Hone}, \citenamefont {Brus},
  \citenamefont {Heinz}, \citenamefont {Reichman},\ and\ \citenamefont
  {Chernikov}}]{Raja:2017aa}%
  \BibitemOpen
  \bibfield  {author} {\bibinfo {author} {\bibfnamefont {Archana}\ \bibnamefont
  {Raja}}, \bibinfo {author} {\bibfnamefont {Andrey}\ \bibnamefont {Chaves}},
  \bibinfo {author} {\bibfnamefont {Jaeeun}\ \bibnamefont {Yu}}, \bibinfo
  {author} {\bibfnamefont {Ghidewon}\ \bibnamefont {Arefe}}, \bibinfo {author}
  {\bibfnamefont {Heather~M.}\ \bibnamefont {Hill}}, \bibinfo {author}
  {\bibfnamefont {Albert~F.}\ \bibnamefont {Rigosi}}, \bibinfo {author}
  {\bibfnamefont {Timothy~C.}\ \bibnamefont {Berkelbach}}, \bibinfo {author}
  {\bibfnamefont {Philipp}\ \bibnamefont {Nagler}}, \bibinfo {author}
  {\bibfnamefont {Christian}\ \bibnamefont {Sch{\"u}ller}}, \bibinfo {author}
  {\bibfnamefont {Tobias}\ \bibnamefont {Korn}}, \bibinfo {author}
  {\bibfnamefont {Colin}\ \bibnamefont {Nuckolls}}, \bibinfo {author}
  {\bibfnamefont {James}\ \bibnamefont {Hone}}, \bibinfo {author}
  {\bibfnamefont {Louis~E.}\ \bibnamefont {Brus}}, \bibinfo {author}
  {\bibfnamefont {Tony~F.}\ \bibnamefont {Heinz}}, \bibinfo {author}
  {\bibfnamefont {David~R.}\ \bibnamefont {Reichman}}, \ and\ \bibinfo {author}
  {\bibfnamefont {Alexey}\ \bibnamefont {Chernikov}},\ }\bibfield  {title}
  {\enquote {\bibinfo {title} {Coulomb engineering of the bandgap and excitons
  in two-dimensional materials},}\ }\href {\doibase 10.1038/ncomms15251}
  {\bibfield  {journal} {\bibinfo  {journal} {Nature Communications}\ }\textbf
  {\bibinfo {volume} {8}},\ \bibinfo {pages} {15251} (\bibinfo {year}
  {2017})}\BibitemShut {NoStop}%
\bibitem [{\citenamefont {Raja}\ \emph {et~al.}(2019)\citenamefont {Raja},
  \citenamefont {Waldecker}, \citenamefont {Zipfel}, \citenamefont {Cho},
  \citenamefont {Brem}, \citenamefont {Ziegler}, \citenamefont {Kulig},
  \citenamefont {Taniguchi}, \citenamefont {Watanabe}, \citenamefont {Malic},
  \citenamefont {Heinz}, \citenamefont {Berkelbach},\ and\ \citenamefont
  {Chernikov}}]{Raja:2019aa}%
  \BibitemOpen
  \bibfield  {author} {\bibinfo {author} {\bibfnamefont {Archana}\ \bibnamefont
  {Raja}}, \bibinfo {author} {\bibfnamefont {Lutz}\ \bibnamefont {Waldecker}},
  \bibinfo {author} {\bibfnamefont {Jonas}\ \bibnamefont {Zipfel}}, \bibinfo
  {author} {\bibfnamefont {Yeongsu}\ \bibnamefont {Cho}}, \bibinfo {author}
  {\bibfnamefont {Samuel}\ \bibnamefont {Brem}}, \bibinfo {author}
  {\bibfnamefont {Jonas~D.}\ \bibnamefont {Ziegler}}, \bibinfo {author}
  {\bibfnamefont {Marvin}\ \bibnamefont {Kulig}}, \bibinfo {author}
  {\bibfnamefont {Takashi}\ \bibnamefont {Taniguchi}}, \bibinfo {author}
  {\bibfnamefont {Kenji}\ \bibnamefont {Watanabe}}, \bibinfo {author}
  {\bibfnamefont {Ermin}\ \bibnamefont {Malic}}, \bibinfo {author}
  {\bibfnamefont {Tony~F.}\ \bibnamefont {Heinz}}, \bibinfo {author}
  {\bibfnamefont {Timothy~C.}\ \bibnamefont {Berkelbach}}, \ and\ \bibinfo
  {author} {\bibfnamefont {Alexey}\ \bibnamefont {Chernikov}},\ }\bibfield
  {title} {\enquote {\bibinfo {title} {Dielectric disorder in two-dimensional
  materials},}\ }\href {\doibase 10.1038/s41565-019-0520-0} {\bibfield
  {journal} {\bibinfo  {journal} {Nature Nanotechnology} {\bf 14}, 832\ } (\bibinfo {year}
  {2019})}\BibitemShut {NoStop}%
\bibitem [{\citenamefont {Robert}\ \emph {et~al.}(2018)\citenamefont {Robert},
  \citenamefont {Semina}, \citenamefont {Cadiz}, \citenamefont {Manca},
  \citenamefont {Courtade}, \citenamefont {Taniguchi}, \citenamefont
  {Watanabe}, \citenamefont {Cai}, \citenamefont {Tongay}, \citenamefont
  {Lassagne}, \citenamefont {Renucci}, \citenamefont {Amand}, \citenamefont
  {Marie}, \citenamefont {Glazov},\ and\ \citenamefont
  {Urbaszek}}]{PhysRevMaterials.2.011001}%
  \BibitemOpen
  \bibfield  {author} {\bibinfo {author} {\bibfnamefont {C.}~\bibnamefont
  {Robert}}, \bibinfo {author} {\bibfnamefont {M.~A.}\ \bibnamefont {Semina}},
  \bibinfo {author} {\bibfnamefont {F.}~\bibnamefont {Cadiz}}, \bibinfo
  {author} {\bibfnamefont {M.}~\bibnamefont {Manca}}, \bibinfo {author}
  {\bibfnamefont {E.}~\bibnamefont {Courtade}}, \bibinfo {author}
  {\bibfnamefont {T.}~\bibnamefont {Taniguchi}}, \bibinfo {author}
  {\bibfnamefont {K.}~\bibnamefont {Watanabe}}, \bibinfo {author}
  {\bibfnamefont {H.}~\bibnamefont {Cai}}, \bibinfo {author} {\bibfnamefont
  {S.}~\bibnamefont {Tongay}}, \bibinfo {author} {\bibfnamefont
  {B.}~\bibnamefont {Lassagne}}, \bibinfo {author} {\bibfnamefont
  {P.}~\bibnamefont {Renucci}}, \bibinfo {author} {\bibfnamefont
  {T.}~\bibnamefont {Amand}}, \bibinfo {author} {\bibfnamefont
  {X.}~\bibnamefont {Marie}}, \bibinfo {author} {\bibfnamefont {M.~M.}\
  \bibnamefont {Glazov}}, \ and\ \bibinfo {author} {\bibfnamefont
  {B.}~\bibnamefont {Urbaszek}},\ }\bibfield  {title} {\enquote {\bibinfo
  {title} {Optical spectroscopy of excited exciton states in
  ${\mathrm{MoS}}_{2}$ monolayers in van der Waals heterostructures},}\ }\href
  {\doibase 10.1103/PhysRevMaterials.2.011001} {\bibfield  {journal} {\bibinfo
  {journal} {Phys. Rev. Materials}\ }\textbf {\bibinfo {volume} {2}},\ \bibinfo
  {pages} {011001} (\bibinfo {year} {2018})}\BibitemShut {NoStop}%
\bibitem [{\citenamefont {Fang}\ \emph {et~al.}(2019)\citenamefont {Fang},
  \citenamefont {Han}, \citenamefont {Robert}, \citenamefont {Semina},
  \citenamefont {Lagarde}, \citenamefont {Courtade}, \citenamefont {Taniguchi},
  \citenamefont {Watanabe}, \citenamefont {Amand}, \citenamefont {Urbaszek},
  \citenamefont {Glazov},\ and\ \citenamefont
  {Marie}}]{PhysRevLett.123.067401}%
  \BibitemOpen
  \bibfield  {author} {\bibinfo {author} {\bibfnamefont {H.~H.}\ \bibnamefont
  {Fang}}, \bibinfo {author} {\bibfnamefont {B.}~\bibnamefont {Han}}, \bibinfo
  {author} {\bibfnamefont {C.}~\bibnamefont {Robert}}, \bibinfo {author}
  {\bibfnamefont {M.~A.}\ \bibnamefont {Semina}}, \bibinfo {author}
  {\bibfnamefont {D.}~\bibnamefont {Lagarde}}, \bibinfo {author} {\bibfnamefont
  {E.}~\bibnamefont {Courtade}}, \bibinfo {author} {\bibfnamefont
  {T.}~\bibnamefont {Taniguchi}}, \bibinfo {author} {\bibfnamefont
  {K.}~\bibnamefont {Watanabe}}, \bibinfo {author} {\bibfnamefont
  {T.}~\bibnamefont {Amand}}, \bibinfo {author} {\bibfnamefont
  {B.}~\bibnamefont {Urbaszek}}, \bibinfo {author} {\bibfnamefont {M.~M.}\
  \bibnamefont {Glazov}}, \ and\ \bibinfo {author} {\bibfnamefont
  {X.}~\bibnamefont {Marie}},\ }\bibfield  {title} {\enquote {\bibinfo {title}
  {Control of the exciton radiative lifetime in van der Waals
  heterostructures},}\ }\href {\doibase 10.1103/PhysRevLett.123.067401}
  {\bibfield  {journal} {\bibinfo  {journal} {Phys. Rev. Lett.}\ }\textbf
  {\bibinfo {volume} {123}},\ \bibinfo {pages} {067401} (\bibinfo {year}
  {2019})}\BibitemShut {NoStop}%
\bibitem [{\citenamefont {Zhou}\ \emph {et~al.}(2020)\citenamefont {Zhou},
  \citenamefont {Scuri}, \citenamefont {Sung}, \citenamefont {Gelly},
  \citenamefont {Wild}, \citenamefont {De~Greve}, \citenamefont {Joe},
  \citenamefont {Taniguchi}, \citenamefont {Watanabe}, \citenamefont {Kim},
  \citenamefont {Lukin},\ and\ \citenamefont {Park}}]{PhysRevLett.124.027401}%
  \BibitemOpen
  \bibfield  {author} {\bibinfo {author} {\bibfnamefont {You}\ \bibnamefont
  {Zhou}}, \bibinfo {author} {\bibfnamefont {Giovanni}\ \bibnamefont {Scuri}},
  \bibinfo {author} {\bibfnamefont {Jiho}\ \bibnamefont {Sung}}, \bibinfo
  {author} {\bibfnamefont {Ryan~J.}\ \bibnamefont {Gelly}}, \bibinfo {author}
  {\bibfnamefont {Dominik~S.}\ \bibnamefont {Wild}}, \bibinfo {author}
  {\bibfnamefont {Kristiaan}\ \bibnamefont {De~Greve}}, \bibinfo {author}
  {\bibfnamefont {Andrew~Y.}\ \bibnamefont {Joe}}, \bibinfo {author}
  {\bibfnamefont {Takashi}\ \bibnamefont {Taniguchi}}, \bibinfo {author}
  {\bibfnamefont {Kenji}\ \bibnamefont {Watanabe}}, \bibinfo {author}
  {\bibfnamefont {Philip}\ \bibnamefont {Kim}}, \bibinfo {author}
  {\bibfnamefont {Mikhail~D.}\ \bibnamefont {Lukin}}, \ and\ \bibinfo {author}
  {\bibfnamefont {Hongkun}\ \bibnamefont {Park}},\ }\bibfield  {title}
  {\enquote {\bibinfo {title} {Controlling excitons in an atomically thin
  membrane with a mirror},}\ }\href {\doibase 10.1103/PhysRevLett.124.027401}
  {\bibfield  {journal} {\bibinfo  {journal} {Phys. Rev. Lett.}\ }\textbf
  {\bibinfo {volume} {124}},\ \bibinfo {pages} {027401} (\bibinfo {year}
  {2020})}\BibitemShut {NoStop}%
\bibitem [{\citenamefont {Rogers}\ \emph {et~al.}(2020)\citenamefont {Rogers},
  \citenamefont {Gray}, \citenamefont {Bogdanowicz}, \citenamefont {Taniguchi},
  \citenamefont {Watanabe},\ and\ \citenamefont
  {Mabuchi}}]{PhysRevResearch.2.012029}%
  \BibitemOpen
  \bibfield  {author} {\bibinfo {author} {\bibfnamefont {Christopher}\
  \bibnamefont {Rogers}}, \bibinfo {author} {\bibfnamefont {Dodd}\ \bibnamefont
  {Gray}}, \bibinfo {author} {\bibfnamefont {Nathan}\ \bibnamefont
  {Bogdanowicz}}, \bibinfo {author} {\bibfnamefont {Takashi}\ \bibnamefont
  {Taniguchi}}, \bibinfo {author} {\bibfnamefont {Kenji}\ \bibnamefont
  {Watanabe}}, \ and\ \bibinfo {author} {\bibfnamefont {Hideo}\ \bibnamefont
  {Mabuchi}},\ }\bibfield  {title} {\enquote {\bibinfo {title} {Coherent
  feedback control of two-dimensional excitons},}\ }\href {\doibase
  10.1103/PhysRevResearch.2.012029} {\bibfield  {journal} {\bibinfo  {journal}
  {Phys. Rev. Research}\ }\textbf {\bibinfo {volume} {2}},\ \bibinfo {pages}
  {012029} (\bibinfo {year} {2020})}\BibitemShut {NoStop}%
%
  \bibitem{Horng:19}
J.~Horng, Y.-H. Chou, T.-C. Chang, C.-Y. Hsu, T.-C. Lu, and H.~Deng, ``Engineering radiative coupling of excitons in 2d semiconductors,''
Optica {\bf 6}, 1443 (2019).
%
\bibitem [{\citenamefont {Baranov}\ \emph {et~al.}(2017)\citenamefont
  {Baranov}, \citenamefont {Krasnok}, \citenamefont {Shegai}, \citenamefont
  {Al{\`u}},\ and\ \citenamefont {Chong}}]{Baranov:2017aa}%
  \BibitemOpen
  \bibfield  {author} {\bibinfo {author} {\bibfnamefont {Denis~G.}\
  \bibnamefont {Baranov}}, \bibinfo {author} {\bibfnamefont {Alex}\
  \bibnamefont {Krasnok}}, \bibinfo {author} {\bibfnamefont {Timur}\
  \bibnamefont {Shegai}}, \bibinfo {author} {\bibfnamefont {Andrea}\
  \bibnamefont {Al{\`u}}}, \ and\ \bibinfo {author} {\bibfnamefont {Yidong}\
  \bibnamefont {Chong}},\ }\bibfield  {title} {\enquote {\bibinfo {title}
  {Coherent perfect absorbers: linear control of light with light},}\ }\href
  {\doibase 10.1038/natrevmats.2017.64} {\bibfield  {journal} {\bibinfo
  {journal} {Nature Reviews Materials}\ }\textbf {\bibinfo {volume} {2}},\
  \bibinfo {pages} {17064} (\bibinfo {year} {2017})}\BibitemShut {NoStop}%
\bibitem [{\citenamefont {Scuri}\ \emph {et~al.}(2018)\citenamefont {Scuri},
  \citenamefont {Zhou}, \citenamefont {High}, \citenamefont {Wild},
  \citenamefont {Shu}, \citenamefont {De~Greve}, \citenamefont {Jauregui},
  \citenamefont {Taniguchi}, \citenamefont {Watanabe}, \citenamefont {Kim},
  \citenamefont {Lukin},\ and\ \citenamefont {Park}}]{PhysRevLett.120.037402}%
  \BibitemOpen
  \bibfield  {author} {\bibinfo {author} {\bibfnamefont {Giovanni}\
  \bibnamefont {Scuri}}, \bibinfo {author} {\bibfnamefont {You}\ \bibnamefont
  {Zhou}}, \bibinfo {author} {\bibfnamefont {Alexander~A.}\ \bibnamefont
  {High}}, \bibinfo {author} {\bibfnamefont {Dominik~S.}\ \bibnamefont {Wild}},
  \bibinfo {author} {\bibfnamefont {Chi}\ \bibnamefont {Shu}}, \bibinfo
  {author} {\bibfnamefont {Kristiaan}\ \bibnamefont {De~Greve}}, \bibinfo
  {author} {\bibfnamefont {Luis~A.}\ \bibnamefont {Jauregui}}, \bibinfo
  {author} {\bibfnamefont {Takashi}\ \bibnamefont {Taniguchi}}, \bibinfo
  {author} {\bibfnamefont {Kenji}\ \bibnamefont {Watanabe}}, \bibinfo {author}
  {\bibfnamefont {Philip}\ \bibnamefont {Kim}}, \bibinfo {author}
  {\bibfnamefont {Mikhail~D.}\ \bibnamefont {Lukin}}, \ and\ \bibinfo {author}
  {\bibfnamefont {Hongkun}\ \bibnamefont {Park}},\ }\bibfield  {title}
  {\enquote {\bibinfo {title} {Large excitonic reflectivity of monolayer
  ${\mathrm{MoSe}}_{2}$ encapsulated in hexagonal boron nitride},}\ }\href
  {\doibase 10.1103/PhysRevLett.120.037402} {\bibfield  {journal} {\bibinfo
  {journal} {Phys. Rev. Lett.}\ }\textbf {\bibinfo {volume} {120}},\ \bibinfo
  {pages} {037402} (\bibinfo {year} {2018})}\BibitemShut {NoStop}%
\bibitem [{\citenamefont {Back}\ \emph {et~al.}(2018)\citenamefont {Back},
  \citenamefont {Zeytinoglu}, \citenamefont {Ijaz}, \citenamefont {Kroner},\
  and\ \citenamefont {Imamo\ifmmode~\breve{g}\else
  \u{g}\fi{}lu}}]{PhysRevLett.120.037401}%
  \BibitemOpen
  \bibfield  {author} {\bibinfo {author} {\bibfnamefont {Patrick}\ \bibnamefont
  {Back}}, \bibinfo {author} {\bibfnamefont {Sina}\ \bibnamefont {Zeytinoglu}},
  \bibinfo {author} {\bibfnamefont {Aroosa}\ \bibnamefont {Ijaz}}, \bibinfo
  {author} {\bibfnamefont {Martin}\ \bibnamefont {Kroner}}, \ and\ \bibinfo
  {author} {\bibfnamefont {Atac}\ \bibnamefont {Imamo\ifmmode~\breve{g}\else
  \u{g}\fi{}lu}},\ }\bibfield  {title} {\enquote {\bibinfo {title} {Realization
  of an electrically tunable narrow-bandwidth atomically thin mirror using
  monolayer ${\mathrm{MoSe}}_{2}$},}\ }\href {\doibase
  10.1103/PhysRevLett.120.037401} {\bibfield  {journal} {\bibinfo  {journal}
  {Phys. Rev. Lett.}\ }\textbf {\bibinfo {volume} {120}},\ \bibinfo {pages}
  {037401} (\bibinfo {year} {2018})}\BibitemShut {NoStop}%
\bibitem [{\citenamefont {Koshelev}\ \emph {et~al.}(2018)\citenamefont
  {Koshelev}, \citenamefont {Sychev}, \citenamefont {Sadrieva}, \citenamefont
  {Bogdanov},\ and\ \citenamefont {Iorsh}}]{PhysRevB.98.161113}%
  \BibitemOpen
  \bibfield  {author} {\bibinfo {author} {\bibfnamefont {K.~L.}\ \bibnamefont
  {Koshelev}}, \bibinfo {author} {\bibfnamefont {S.~K.}\ \bibnamefont
  {Sychev}}, \bibinfo {author} {\bibfnamefont {Z.~F.}\ \bibnamefont
  {Sadrieva}}, \bibinfo {author} {\bibfnamefont {A.~A.}\ \bibnamefont
  {Bogdanov}}, \ and\ \bibinfo {author} {\bibfnamefont {I.~V.}\ \bibnamefont
  {Iorsh}},\ }\bibfield  {title} {\enquote {\bibinfo {title} {Strong coupling
  between excitons in transition metal dichalcogenides and optical bound states
  in the continuum},}\ }\href {\doibase 10.1103/PhysRevB.98.161113} {\bibfield
  {journal} {\bibinfo  {journal} {Phys. Rev. B}\ }\textbf {\bibinfo {volume}
  {98}},\ \bibinfo {pages} {161113} (\bibinfo {year} {2018})}\BibitemShut
  {NoStop}%
\bibitem [{\citenamefont {Krasnok}\ \emph {et~al.}(2018)\citenamefont
  {Krasnok}, \citenamefont {Lepeshov},\ and\ \citenamefont
  {Al\'{u}}}]{Krasnok:18}%
  \BibitemOpen
  \bibfield  {author} {\bibinfo {author} {\bibfnamefont {Alex}\ \bibnamefont
  {Krasnok}}, \bibinfo {author} {\bibfnamefont {Sergey}\ \bibnamefont
  {Lepeshov}}, \ and\ \bibinfo {author} {\bibfnamefont {Andrea}\ \bibnamefont
  {Al\'{u}}},\ }\bibfield  {title} {\enquote {\bibinfo {title} {Nanophotonics
  with 2d transition metal dichalcogenides},}\ }\href {\doibase
  10.1364/OE.26.015972} {\bibfield  {journal} {\bibinfo  {journal} {Opt.
  Express}\ }\textbf {\bibinfo {volume} {26}},\ \bibinfo {pages} {15972--15994}
  (\bibinfo {year} {2018})}\BibitemShut {NoStop}%
\bibitem [{\citenamefont {Xu}\ \emph {et~al.}(2014)\citenamefont {Xu},
  \citenamefont {Yao}, \citenamefont {Xiao},\ and\ \citenamefont
  {Heinz}}]{Xu:2014cr}%
  \BibitemOpen
  \bibfield  {author} {\bibinfo {author} {\bibfnamefont {Xiaodong}\
  \bibnamefont {Xu}}, \bibinfo {author} {\bibfnamefont {Wang}\ \bibnamefont
  {Yao}}, \bibinfo {author} {\bibfnamefont {Di}~\bibnamefont {Xiao}}, \ and\
  \bibinfo {author} {\bibfnamefont {Tony~F.}\ \bibnamefont {Heinz}},\
  }\bibfield  {title} {\enquote {\bibinfo {title} {Spin and pseudospins in
  layered transition metal dichalcogenides},}\ }\href
  {http://dx.doi.org/10.1038/nphys2942} {\bibfield  {journal} {\bibinfo
  {journal} {Nat Phys}\ }\textbf {\bibinfo {volume} {10}},\ \bibinfo {pages}
  {343--350} (\bibinfo {year} {2014})}\BibitemShut {NoStop}%
\bibitem [{\citenamefont {Xiao}\ \emph {et~al.}(2012)\citenamefont {Xiao},
  \citenamefont {Liu}, \citenamefont {Feng}, \citenamefont {Xu},\ and\
  \citenamefont {Yao}}]{Xiao:2012cr}%
  \BibitemOpen
  \bibfield  {author} {\bibinfo {author} {\bibfnamefont {Di}~\bibnamefont
  {Xiao}}, \bibinfo {author} {\bibfnamefont {Gui-Bin}\ \bibnamefont {Liu}},
  \bibinfo {author} {\bibfnamefont {Wanxiang}\ \bibnamefont {Feng}}, \bibinfo
  {author} {\bibfnamefont {Xiaodong}\ \bibnamefont {Xu}}, \ and\ \bibinfo
  {author} {\bibfnamefont {Wang}\ \bibnamefont {Yao}},\ }\bibfield  {title}
  {\enquote {\bibinfo {title} {Coupled spin and valley physics in monolayers of
  {MoS}$_{2}$ and other group-{VI} dichalcogenides},}\ }\href {\doibase
  10.1103/PhysRevLett.108.196802} {\bibfield  {journal} {\bibinfo  {journal}
  {Phys. Rev. Lett.}\ }\textbf {\bibinfo {volume} {108}},\ \bibinfo {pages}
  {196802} (\bibinfo {year} {2012})}\BibitemShut {NoStop}%
\bibitem [{\citenamefont {Mak}\ \emph {et~al.}(2012)\citenamefont {Mak},
  \citenamefont {He}, \citenamefont {Shan},\ and\ \citenamefont
  {Heinz}}]{Mak:2012qf}%
  \BibitemOpen
  \bibfield  {author} {\bibinfo {author} {\bibfnamefont {Kin~Fai}\ \bibnamefont
  {Mak}}, \bibinfo {author} {\bibfnamefont {Keliang}\ \bibnamefont {He}},
  \bibinfo {author} {\bibfnamefont {Jie}\ \bibnamefont {Shan}}, \ and\ \bibinfo
  {author} {\bibfnamefont {Tony~F.}\ \bibnamefont {Heinz}},\ }\bibfield
  {title} {\enquote {\bibinfo {title} {Control of valley polarization in
  monolayer MoS$_2$ by optical helicity},}\ }\href
  {http://dx.doi.org/10.1038/nnano.2012.96} {\bibfield  {journal} {\bibinfo
  {journal} {Nat Nano}\ }\textbf {\bibinfo {volume} {7}},\ \bibinfo {pages}
  {494--498} (\bibinfo {year} {2012})}\BibitemShut {NoStop}%
\bibitem [{\citenamefont {Sallen}\ \emph {et~al.}(2012)\citenamefont {Sallen},
  \citenamefont {Bouet}, \citenamefont {Marie}, \citenamefont {Wang},
  \citenamefont {Zhu}, \citenamefont {Han}, \citenamefont {Lu}, \citenamefont
  {Tan}, \citenamefont {Amand}, \citenamefont {Liu},\ and\ \citenamefont
  {Urbaszek}}]{Sallen:2012qf}%
  \BibitemOpen
  \bibfield  {author} {\bibinfo {author} {\bibfnamefont {G.}~\bibnamefont
  {Sallen}}, \bibinfo {author} {\bibfnamefont {L.}~\bibnamefont {Bouet}},
  \bibinfo {author} {\bibfnamefont {X.}~\bibnamefont {Marie}}, \bibinfo
  {author} {\bibfnamefont {G.}~\bibnamefont {Wang}}, \bibinfo {author}
  {\bibfnamefont {C.~R.}\ \bibnamefont {Zhu}}, \bibinfo {author} {\bibfnamefont
  {W.~P.}\ \bibnamefont {Han}}, \bibinfo {author} {\bibfnamefont
  {Y.}~\bibnamefont {Lu}}, \bibinfo {author} {\bibfnamefont {P.~H.}\
  \bibnamefont {Tan}}, \bibinfo {author} {\bibfnamefont {T.}~\bibnamefont
  {Amand}}, \bibinfo {author} {\bibfnamefont {B.~L.}\ \bibnamefont {Liu}}, \
  and\ \bibinfo {author} {\bibfnamefont {B.}~\bibnamefont {Urbaszek}},\
  }\bibfield  {title} {\enquote {\bibinfo {title} {Robust optical emission
  polarization in MoS${}_{2}$ monolayers through selective valley
  excitation},}\ }\href {\doibase 10.1103/PhysRevB.86.081301} {\bibfield
  {journal} {\bibinfo  {journal} {Phys. Rev. B}\ }\textbf {\bibinfo {volume}
  {86}},\ \bibinfo {pages} {081301} (\bibinfo {year} {2012})}\BibitemShut
  {NoStop}%
\bibitem [{\citenamefont {Kioseoglou}\ \emph {et~al.}(2012)\citenamefont
  {Kioseoglou}, \citenamefont {Hanbicki}, \citenamefont {Currie}, \citenamefont
  {Friedman}, \citenamefont {Gunlycke},\ and\ \citenamefont
  {Jonker}}]{Kioseoglou}%
  \BibitemOpen
  \bibfield  {author} {\bibinfo {author} {\bibfnamefont {G.}~\bibnamefont
  {Kioseoglou}}, \bibinfo {author} {\bibfnamefont {A.~T.}\ \bibnamefont
  {Hanbicki}}, \bibinfo {author} {\bibfnamefont {M.}~\bibnamefont {Currie}},
  \bibinfo {author} {\bibfnamefont {A.~L.}\ \bibnamefont {Friedman}}, \bibinfo
  {author} {\bibfnamefont {D.}~\bibnamefont {Gunlycke}}, \ and\ \bibinfo
  {author} {\bibfnamefont {B.~T.}\ \bibnamefont {Jonker}},\ }\bibfield  {title}
  {\enquote {\bibinfo {title} {Valley polarization and intervalley scattering
  in monolayer MoS$_2$},}\ }\href {\doibase
  http://dx.doi.org/10.1063/1.4768299} {\bibfield  {journal} {\bibinfo
  {journal} {Applied Physics Letters}\ }\textbf {\bibinfo {volume} {101}},\
  \bibinfo {eid} {221907} (\bibinfo {year} {2012})}\BibitemShut {NoStop}%
\bibitem [{\citenamefont {Zeng}\ \emph {et~al.}(2012)\citenamefont {Zeng},
  \citenamefont {Dai}, \citenamefont {Yao}, \citenamefont {Xiao},\ and\
  \citenamefont {Cui}}]{Zeng:2012ys}%
  \BibitemOpen
  \bibfield  {author} {\bibinfo {author} {\bibfnamefont {Hualing}\ \bibnamefont
  {Zeng}}, \bibinfo {author} {\bibfnamefont {Junfeng}\ \bibnamefont {Dai}},
  \bibinfo {author} {\bibfnamefont {Wang}\ \bibnamefont {Yao}}, \bibinfo
  {author} {\bibfnamefont {Di}~\bibnamefont {Xiao}}, \ and\ \bibinfo {author}
  {\bibfnamefont {Xiaodong}\ \bibnamefont {Cui}},\ }\bibfield  {title}
  {\enquote {\bibinfo {title} {Valley polarization in MoS$_2$ monolayers by
  optical pumping},}\ }\href {http://dx.doi.org/10.1038/nnano.2012.95}
  {\bibfield  {journal} {\bibinfo  {journal} {Nat Nano}\ }\textbf {\bibinfo
  {volume} {7}},\ \bibinfo {pages} {490--493} (\bibinfo {year}
  {2012})}\BibitemShut {NoStop}%
\bibitem [{\citenamefont {Jones}\ \emph {et~al.}(2013)\citenamefont {Jones},
  \citenamefont {Yu}, \citenamefont {Ghimire}, \citenamefont {Wu},
  \citenamefont {Aivazian}, \citenamefont {Ross}, \citenamefont {Zhao},
  \citenamefont {Yan}, \citenamefont {Mandrus}, \citenamefont {Xiao},
  \citenamefont {Yao},\ and\ \citenamefont {Xu}}]{Jones:2013tg}%
  \BibitemOpen
  \bibfield  {author} {\bibinfo {author} {\bibfnamefont {Aaron~M.}\
  \bibnamefont {Jones}}, \bibinfo {author} {\bibfnamefont {Hongyi}\
  \bibnamefont {Yu}}, \bibinfo {author} {\bibfnamefont {Nirmal~J.}\
  \bibnamefont {Ghimire}}, \bibinfo {author} {\bibfnamefont {Sanfeng}\
  \bibnamefont {Wu}}, \bibinfo {author} {\bibfnamefont {Grant}\ \bibnamefont
  {Aivazian}}, \bibinfo {author} {\bibfnamefont {Jason~S.}\ \bibnamefont
  {Ross}}, \bibinfo {author} {\bibfnamefont {Bo}~\bibnamefont {Zhao}}, \bibinfo
  {author} {\bibfnamefont {Jiaqiang}\ \bibnamefont {Yan}}, \bibinfo {author}
  {\bibfnamefont {David~G.}\ \bibnamefont {Mandrus}}, \bibinfo {author}
  {\bibfnamefont {Di}~\bibnamefont {Xiao}}, \bibinfo {author} {\bibfnamefont
  {Wang}\ \bibnamefont {Yao}}, \ and\ \bibinfo {author} {\bibfnamefont
  {Xiaodong}\ \bibnamefont {Xu}},\ }\bibfield  {title} {\enquote {\bibinfo
  {title} {Optical generation of excitonic valley coherence in monolayer
  WSe$_2$},}\ }\href {http://dx.doi.org/10.1038/nnano.2013.151} {\bibfield
  {journal} {\bibinfo  {journal} {Nat Nano}\ }\textbf {\bibinfo {volume} {8}},\
  \bibinfo {pages} {634--638} (\bibinfo {year} {2013})}\BibitemShut {NoStop}%
\bibitem [{\citenamefont {Yu}\ \emph {et~al.}(2014)\citenamefont {Yu},
  \citenamefont {Liu}, \citenamefont {Gong}, \citenamefont {Xu},\ and\
  \citenamefont {Yao}}]{Yu:2014fk-1}%
  \BibitemOpen
  \bibfield  {author} {\bibinfo {author} {\bibfnamefont {Hongyi}\ \bibnamefont
  {Yu}}, \bibinfo {author} {\bibfnamefont {Gui-Bin}\ \bibnamefont {Liu}},
  \bibinfo {author} {\bibfnamefont {Pu}~\bibnamefont {Gong}}, \bibinfo {author}
  {\bibfnamefont {Xiaodong}\ \bibnamefont {Xu}}, \ and\ \bibinfo {author}
  {\bibfnamefont {Wang}\ \bibnamefont {Yao}},\ }\bibfield  {title} {\enquote
  {\bibinfo {title} {Dirac cones and {D}irac saddle points of bright excitons
  in monolayer transition metal dichalcogenides},}\ }\href
  {http://dx.doi.org/10.1038/ncomms4876} {\bibfield  {journal} {\bibinfo
  {journal} {Nat Commun}\ }\textbf {\bibinfo {volume} {5}},\ \bibinfo {pages}
  {3876} (\bibinfo {year} {2014})}\BibitemShut {NoStop}%
\bibitem [{\citenamefont {Selig}\ \emph {et~al.}(2019)\citenamefont {Selig},
  \citenamefont {Katsch}, \citenamefont {Schmidt}, \citenamefont {Michaelis~de
  Vasconcellos}, \citenamefont {Bratschitsch}, \citenamefont {Malic},\ and\
  \citenamefont {Knorr}}]{PhysRevResearch.1.022007}%
  \BibitemOpen
  \bibfield  {author} {\bibinfo {author} {\bibfnamefont {Malte}\ \bibnamefont
  {Selig}}, \bibinfo {author} {\bibfnamefont {Florian}\ \bibnamefont {Katsch}},
  \bibinfo {author} {\bibfnamefont {Robert}\ \bibnamefont {Schmidt}}, \bibinfo
  {author} {\bibfnamefont {Steffen}\ \bibnamefont {Michaelis~de Vasconcellos}},
  \bibinfo {author} {\bibfnamefont {Rudolf}\ \bibnamefont {Bratschitsch}},
  \bibinfo {author} {\bibfnamefont {Ermin}\ \bibnamefont {Malic}}, \ and\
  \bibinfo {author} {\bibfnamefont {Andreas}\ \bibnamefont {Knorr}},\
  }\bibfield  {title} {\enquote {\bibinfo {title} {Ultrafast dynamics in
  monolayer transition metal dichalcogenides: Interplay of dark excitons,
  phonons, and intervalley exchange},}\ }\href {\doibase
  10.1103/PhysRevResearch.1.022007} {\bibfield  {journal} {\bibinfo  {journal}
  {Phys. Rev. Research}\ }\textbf {\bibinfo {volume} {1}},\ \bibinfo {pages}
  {022007} (\bibinfo {year} {2019})}\BibitemShut {NoStop}%
\bibitem [{\citenamefont {Shinokita}\ \emph {et~al.}(2019)\citenamefont
  {Shinokita}, \citenamefont {Wang}, \citenamefont {Miyauchi}, \citenamefont
  {Watanabe}, \citenamefont {Taniguchi},\ and\ \citenamefont
  {Matsuda}}]{Shinokita:2019aa}%
  \BibitemOpen
  \bibfield  {author} {\bibinfo {author} {\bibfnamefont {Keisuke}\ \bibnamefont
  {Shinokita}}, \bibinfo {author} {\bibfnamefont {Xiaofan}\ \bibnamefont
  {Wang}}, \bibinfo {author} {\bibfnamefont {Yuhei}\ \bibnamefont {Miyauchi}},
  \bibinfo {author} {\bibfnamefont {Kenji}\ \bibnamefont {Watanabe}}, \bibinfo
  {author} {\bibfnamefont {Takashi}\ \bibnamefont {Taniguchi}}, \ and\ \bibinfo
  {author} {\bibfnamefont {Kazunari}\ \bibnamefont {Matsuda}},\ }\bibfield
  {title} {\enquote {\bibinfo {title} {Continuous control and enhancement of
  excitonic valley polarization in monolayer WSe$_2$ by electrostatic doping},}\
  }\bibfield  {booktitle} \href {\doibase 10.1002/adfm.201900260} {\bibfield  {journal}
  {\bibinfo  {journal} {Advanced Functional Materials}\ }\textbf {\bibinfo
  {volume} {29}},\ \bibinfo {pages} {1900260} (\bibinfo {year}
  {2019})}\BibitemShut {NoStop}%
\bibitem [{\citenamefont {Tang}\ \emph {et~al.}(2019)\citenamefont {Tang},
  \citenamefont {Mak},\ and\ \citenamefont {Shan}}]{Tang:2019ab}%
  \BibitemOpen
  \bibfield  {author} {\bibinfo {author} {\bibfnamefont {Yanhao}\ \bibnamefont
  {Tang}}, \bibinfo {author} {\bibfnamefont {Kin~Fai}\ \bibnamefont {Mak}}, \
  and\ \bibinfo {author} {\bibfnamefont {Jie}\ \bibnamefont {Shan}},\
  }\bibfield  {title} {\enquote {\bibinfo {title} {Long valley lifetime of dark
  excitons in single-layer WSe$_2$},}\ }\href {\doibase
  10.1038/s41467-019-12129-1} {\bibfield  {journal} {\bibinfo  {journal}
  {Nature Communications}\ }\textbf {\bibinfo {volume} {10}},\ \bibinfo {pages}
  {4047} (\bibinfo {year} {2019})}\BibitemShut {NoStop}%
\bibitem{Dufferwiel:2017aa}
S.~Dufferwiel, T.~P. Lyons, D.~D. Solnyshkov, A.~A.~P. Trichet, F.~Withers,
  S.~Schwarz, G.~Malpuech, J.~M. Smith, K.~S. Novoselov, M.~S. Skolnick, D.~N.
  Krizhanovskii, and A.~I. Tartakovskii, ``Valley-addressable polaritons in atomically thin semiconductors,'' Nature Photonics {\bf 11}, 497 (2017).
\bibitem [{\citenamefont {Yang}\ \emph {et~al.}(2020)\citenamefont {Yang},
  \citenamefont {Robert}, \citenamefont {Lu}, \citenamefont {Van~Tuan},
  \citenamefont {Smirnov}, \citenamefont {Marie},\ and\ \citenamefont
  {Dery}}]{PhysRevB.101.115307}%
  \BibitemOpen
  \bibfield  {author} {\bibinfo {author} {\bibfnamefont {Min}\ \bibnamefont
  {Yang}}, \bibinfo {author} {\bibfnamefont {Cedric}\ \bibnamefont {Robert}},
  \bibinfo {author} {\bibfnamefont {Zhengguang}\ \bibnamefont {Lu}}, \bibinfo
  {author} {\bibfnamefont {Dinh}\ \bibnamefont {Van~Tuan}}, \bibinfo {author}
  {\bibfnamefont {Dmitry}\ \bibnamefont {Smirnov}}, \bibinfo {author}
  {\bibfnamefont {Xavier}\ \bibnamefont {Marie}}, \ and\ \bibinfo {author}
  {\bibfnamefont {Hanan}\ \bibnamefont {Dery}},\ }\bibfield  {title} {\enquote
  {\bibinfo {title} {Exciton valley depolarization in monolayer
  transition-metal dichalcogenides},}\ }\href {\doibase
  10.1103/PhysRevB.101.115307} {\bibfield  {journal} {\bibinfo  {journal}
  {Phys. Rev. B}\ }\textbf {\bibinfo {volume} {101}},\ \bibinfo {pages}
  {115307} (\bibinfo {year} {2020})}\BibitemShut {NoStop}%
%  
\bibitem{PhysRevResearch.2.023322}
M.~Selig, F.~Katsch, S.~Brem, G.~F. Mkrtchian, E.~Malic, and A.~Knorr, ``Suppression of intervalley exchange coupling in the presence of
  momentum-dark states in transition metal dichalcogenides,'' \href{https://doi.org/10.1103/PhysRevResearch.2.023322}{Phys. Rev. Research {\bf 2}, 023322 (2020)}.
%  
\bibitem [{\citenamefont {Glazov}\ \emph {et~al.}(2014)\citenamefont {Glazov},
  \citenamefont {Amand}, \citenamefont {Marie}, \citenamefont {Lagarde},
  \citenamefont {Bouet},\ and\ \citenamefont {Urbaszek}}]{glazov2014exciton}%
  \BibitemOpen
  \bibfield  {author} {\bibinfo {author} {\bibfnamefont {M.~M.}\ \bibnamefont
  {Glazov}}, \bibinfo {author} {\bibfnamefont {T.}~\bibnamefont {Amand}},
  \bibinfo {author} {\bibfnamefont {X.}~\bibnamefont {Marie}}, \bibinfo
  {author} {\bibfnamefont {D.}~\bibnamefont {Lagarde}}, \bibinfo {author}
  {\bibfnamefont {L.}~\bibnamefont {Bouet}}, \ and\ \bibinfo {author}
  {\bibfnamefont {B.}~\bibnamefont {Urbaszek}},\ }\bibfield  {title} {\enquote
  {\bibinfo {title} {Exciton fine structure and spin decoherence in monolayers
  of transition metal dichalcogenides},}\ }\href {\doibase
  10.1103/physrevb.89.201302} {\bibfield  {journal} {\bibinfo  {journal} {Phys.
  Rev. B}\ }\textbf {\bibinfo {volume} {89}},\ \bibinfo {pages} {201302}
  (\bibinfo {year} {2014})}\BibitemShut {NoStop}%
\bibitem [{\citenamefont {Glazov}\ \emph {et~al.}(2015)\citenamefont {Glazov},
  \citenamefont {Ivchenko}, \citenamefont {Wang}, \citenamefont {Amand},
  \citenamefont {Marie}, \citenamefont {Urbaszek},\ and\ \citenamefont
  {Liu}}]{PSSB:PSSB201552211}%
  \BibitemOpen
  \bibfield  {author} {\bibinfo {author} {\bibfnamefont {M.~M.}\ \bibnamefont
  {Glazov}}, \bibinfo {author} {\bibfnamefont {E.~L.}\ \bibnamefont
  {Ivchenko}}, \bibinfo {author} {\bibfnamefont {G.}~\bibnamefont {Wang}},
  \bibinfo {author} {\bibfnamefont {T.}~\bibnamefont {Amand}}, \bibinfo
  {author} {\bibfnamefont {X.}~\bibnamefont {Marie}}, \bibinfo {author}
  {\bibfnamefont {B.}~\bibnamefont {Urbaszek}}, \ and\ \bibinfo {author}
  {\bibfnamefont {B.~L.}\ \bibnamefont {Liu}},\ }\bibfield  {title} {\enquote
  {\bibinfo {title} {Spin and valley dynamics of excitons in transition metal
  dichalcogenide monolayers},}\ }\href {\doibase 10.1002/pssb.201552211}
  {\bibfield  {journal} {\bibinfo  {journal} {physica status solidi (b)}\
  }\textbf {\bibinfo {volume} {252}},\ \bibinfo {pages} {2349--2362} (\bibinfo
  {year} {2015})}\BibitemShut {NoStop}%
\bibitem [{\citenamefont {Yu}\ and\ \citenamefont
  {Wu}(2016)}]{PhysRevB.93.045414}%
  \BibitemOpen
  \bibfield  {author} {\bibinfo {author} {\bibfnamefont {T.}~\bibnamefont
  {Yu}}\ and\ \bibinfo {author} {\bibfnamefont {M.~W.}\ \bibnamefont {Wu}},\
  }\bibfield  {title} {\enquote {\bibinfo {title} {Valley depolarization
  dynamics and valley Hall effect of excitons in monolayer and bilayer
  ${\mathrm{MoS}}_{2}$},}\ }\href {\doibase 10.1103/PhysRevB.93.045414}
  {\bibfield  {journal} {\bibinfo  {journal} {Phys. Rev. B}\ }\textbf {\bibinfo
  {volume} {93}},\ \bibinfo {pages} {045414} (\bibinfo {year}
  {2016})}\BibitemShut {NoStop}%
\bibitem [{\citenamefont {Maialle}\ \emph {et~al.}(1993)\citenamefont
  {Maialle}, \citenamefont {de~Andrada~e Silva},\ and\ \citenamefont
  {Sham}}]{maialle93}%
  \BibitemOpen
  \bibfield  {author} {\bibinfo {author} {\bibfnamefont {M.Z.}\ \bibnamefont
  {Maialle}}, \bibinfo {author} {\bibfnamefont {E.A.}\ \bibnamefont
  {de~Andrada~e Silva}}, \ and\ \bibinfo {author} {\bibfnamefont {L.J.}\
  \bibnamefont {Sham}},\ }\bibfield  {title} {\enquote {\bibinfo {title}
  {Exciton spin dynamics in quantum wells},}\ }\href@noop {} {\bibfield
  {journal} {\bibinfo  {journal} {Phys. Rev. B}\ }\textbf {\bibinfo {volume}
  {47}},\ \bibinfo {pages} {15776} (\bibinfo {year} {1993})}\BibitemShut
  {NoStop}%
\bibitem [{\citenamefont {Goupalov}\ \emph {et~al.}(1998)\citenamefont
  {Goupalov}, \citenamefont {Ivchenko},\ and\ \citenamefont
  {Kavokin}}]{goupalov98}%
  \BibitemOpen
  \bibfield  {author} {\bibinfo {author} {\bibfnamefont {S.~V.}\ \bibnamefont
  {Goupalov}}, \bibinfo {author} {\bibfnamefont {E.~L.}\ \bibnamefont
  {Ivchenko}}, \ and\ \bibinfo {author} {\bibfnamefont {A.~V.}\ \bibnamefont
  {Kavokin}},\ }\bibfield  {title} {\enquote {\bibinfo {title} {Fine structure
  of localized exciton levels in quantum wells},}\ }\href@noop {} {\bibfield
  {journal} {\bibinfo  {journal} {JETP}\ }\textbf {\bibinfo {volume} {86}},\
  \bibinfo {pages} {388} (\bibinfo {year} {1998})}\BibitemShut {NoStop}%
\bibitem [{\citenamefont {Ivchenko}(2005)}]{ivchenko05a}%
  \BibitemOpen
  \bibfield  {author} {\bibinfo {author} {\bibfnamefont {E.~L.}\ \bibnamefont
  {Ivchenko}},\ }\href@noop {} {\emph {\bibinfo {title} {Optical spectroscopy
  of semiconductor nanostructures}}}\ (\bibinfo  {publisher} {Alpha Science,
  Harrow UK},\ \bibinfo {year} {2005})\BibitemShut {NoStop}%
\bibitem [{\citenamefont {Pikus}\ and\ \citenamefont {Bir}(1971)}]{BP_exch71}%
  \BibitemOpen
  \bibfield  {author} {\bibinfo {author} {\bibfnamefont {G.~E.}\ \bibnamefont
  {Pikus}}\ and\ \bibinfo {author} {\bibfnamefont {G.~L.}\ \bibnamefont
  {Bir}},\ }\bibfield  {title} {\enquote {\bibinfo {title} {Exchange
  interaction in excitons in semiconductors},}\ }\href@noop {} {\bibfield
  {journal} {\bibinfo  {journal} {JETP}\ }\textbf {\bibinfo {volume} {33}},\
  \bibinfo {pages} {108} (\bibinfo {year} {1971})}\BibitemShut {NoStop}%
\bibitem [{\citenamefont {Denisov}\ and\ \citenamefont
  {Makarov}(1973)}]{denisovmakarov}%
  \BibitemOpen
  \bibfield  {author} {\bibinfo {author} {\bibfnamefont {M.~M.}\ \bibnamefont
  {Denisov}}\ and\ \bibinfo {author} {\bibfnamefont {V.~P.}\ \bibnamefont
  {Makarov}},\ }\bibfield  {title} {\enquote {\bibinfo {title} {Longitudinal
  and transverse excitons in semiconductors},}\ }\href
  {http://dx.doi.org/10.1002/pssb.2220560102} {\bibfield  {journal} {\bibinfo
  {journal} {Physica Status Solidi (b)}\ }\textbf {\bibinfo {volume} {56}},\
  \bibinfo {pages} {9--59} (\bibinfo {year} {1973})}\BibitemShut {NoStop}%
\bibitem [{\citenamefont {Bir}\ and\ \citenamefont
  {Pikus}(1974)}]{birpikus_eng}%
  \BibitemOpen
  \bibfield  {author} {\bibinfo {author} {\bibfnamefont {G.~L.}\ \bibnamefont
  {Bir}}\ and\ \bibinfo {author} {\bibfnamefont {G.~E.}\ \bibnamefont
  {Pikus}},\ }\href@noop {} {\emph {\bibinfo {title} {Symmetry and
  Strain-induced Effects in Semiconductors}}}\ (\bibinfo  {publisher}
  {Wiley/Halsted Press},\ \bibinfo {year} {1974})\BibitemShut {NoStop}%
 %
 \bibitem{PhysRevX.7.021026}
F.~Cadiz, E.~Courtade, C.~Robert, G.~Wang, Y.~Shen, H.~Cai, T.~Taniguchi,
  K.~Watanabe, H.~Carrere, D.~Lagarde, M.~Manca, T.~Amand, P.~Renucci,
  S.~Tongay, X.~Marie, and B.~Urbaszek, ``Excitonic linewidth approaching the homogeneous limit in
  MoS$_{2}$-based van der Waals heterostructures,'' Phys. Rev. X {\bf 7}, 021026 (2017).
%
\bibitem{Ajayi_2017}
O.~A. Ajayi, J.~V. Ardelean, G.~D. Shepard, J.~Wang, A.~Antony, T.~Taniguchi,
  K.~Watanabe, T.~F. Heinz, S.~Strauf, X.-Y. Zhu, and J.~C. Hone, ``Approaching the intrinsic photoluminescence linewidth in transition
  metal dichalcogenide monolayers,''
2D Materials {\bf 4}, 031011 (2017).
 %
\bibitem [{\citenamefont {Cho}\ and\ \citenamefont
  {Berkelbach}(2018)}]{PhysRevB.97.041409}%
  \BibitemOpen
  \bibfield  {author} {\bibinfo {author} {\bibfnamefont {Yeongsu}\ \bibnamefont
  {Cho}}\ and\ \bibinfo {author} {\bibfnamefont {Timothy~C.}\ \bibnamefont
  {Berkelbach}},\ }\bibfield  {title} {\enquote {\bibinfo {title}
  {Environmentally sensitive theory of electronic and optical transitions in
  atomically thin semiconductors},}\ }\href {\doibase
  10.1103/PhysRevB.97.041409} {\bibfield  {journal} {\bibinfo  {journal} {Phys.
  Rev. B}\ }\textbf {\bibinfo {volume} {97}},\ \bibinfo {pages} {041409}
  (\bibinfo {year} {2018})}\BibitemShut {NoStop}%
%
\bibitem{PhysRevB.98.245126}
I.~C. Gerber and X.~Marie, ``Dependence of band structure and exciton properties of encapsulated
  WSe$_{2}$ monolayers on the hBN-layer thickness,'' Phys. Rev. B {\bf 98}, 245126 (2018).
%
\bibitem [{\citenamefont {Kiselev}\ and\ \citenamefont
  {Zhilich}(1974{\natexlab{a}})}]{zhilich74:eng}%
  \BibitemOpen
  \bibfield  {author} {\bibinfo {author} {\bibfnamefont {V.~A.}\ \bibnamefont
  {Kiselev}}\ and\ \bibinfo {author} {\bibfnamefont {A.~G.}\ \bibnamefont
  {Zhilich}},\ }\bibfield  {title} {\enquote {\bibinfo {title} {On dielectric
  theory of semiconductors with account for excitons},}\ }\href@noop {}
  {\bibfield  {journal} {\bibinfo  {journal} {Sov. Phys. - Semicond.}\ }\textbf
  {\bibinfo {volume} {8}},\ \bibinfo {pages} {411} (\bibinfo {year}
  {1974}{\natexlab{a}})}\BibitemShut {NoStop}%
\bibitem [{\citenamefont {Kiselev}\ and\ \citenamefont
  {Zhilich}(1974{\natexlab{b}})}]{Kiselev74}%
  \BibitemOpen
  \bibfield  {author} {\bibinfo {author} {\bibfnamefont {V.~A.}\ \bibnamefont
  {Kiselev}}\ and\ \bibinfo {author} {\bibfnamefont {A.~G.}\ \bibnamefont
  {Zhilich}},\ }\bibfield  {title} {\enquote {\bibinfo {title} {Effective
  screening of the short-range exchange interaction in excitons},}\ }\href@noop
  {} {\bibfield  {journal} {\bibinfo  {journal} {Phys. Solid. State}\ }\textbf
  {\bibinfo {volume} {15}},\ \bibinfo {pages} {1351} (\bibinfo {year}
  {1974}{\natexlab{b}})}\BibitemShut {NoStop}%
\bibitem [{\citenamefont {Landau}\ and\ \citenamefont
  {Lifshitz}(2004)}]{ll8_eng}%
  \BibitemOpen
  \bibfield  {author} {\bibinfo {author} {\bibfnamefont {L.D.}\ \bibnamefont
  {Landau}}\ and\ \bibinfo {author} {\bibfnamefont {E.M.}\ \bibnamefont
  {Lifshitz}},\ }\href@noop {} {{\selectlanguage {english}\emph {\bibinfo
  {title} {Electrodynamics of Continuous Media (vol. 8)}}}}\ (\bibinfo
  {publisher} {Butterworth-Heinemann, Oxford},\ \bibinfo {year}
  {2004})\BibitemShut {NoStop}%
\bibitem [{\citenamefont {Dyakonov}\ and\ \citenamefont
  {Perel'}(1972)}]{dyakonov72}%
  \BibitemOpen
  \bibfield  {author} {\bibinfo {author} {\bibfnamefont {M.I.}\ \bibnamefont
  {Dyakonov}}\ and\ \bibinfo {author} {\bibfnamefont {V.I}\ \bibnamefont
  {Perel'}},\ }\bibfield  {title} {\enquote {\bibinfo {title} {Spin relaxation
  of conduction electrons in noncentrosymmetric semiconductors},}\ }\href@noop
  {} {\bibfield  {journal} {\bibinfo  {journal} {Sov. Phys. Solid State}\
  }\textbf {\bibinfo {volume} {13}},\ \bibinfo {pages} {3023} (\bibinfo {year}
  {1972})}\BibitemShut {NoStop}%
\bibitem [{\citenamefont {Selig}\ \emph {et~al.}(2018)\citenamefont {Selig},
  \citenamefont {Bergh{\"a}user}, \citenamefont {Richter}, \citenamefont
  {Bratschitsch}, \citenamefont {Knorr},\ and\ \citenamefont
  {Malic}}]{Selig_2018}%
  \BibitemOpen
  \bibfield  {author} {\bibinfo {author} {\bibfnamefont {Malte}\ \bibnamefont
  {Selig}}, \bibinfo {author} {\bibfnamefont {Gunnar}\ \bibnamefont
  {Bergh{\"a}user}}, \bibinfo {author} {\bibfnamefont {Marten}\ \bibnamefont
  {Richter}}, \bibinfo {author} {\bibfnamefont {Rudolf}\ \bibnamefont
  {Bratschitsch}}, \bibinfo {author} {\bibfnamefont {Andreas}\ \bibnamefont
  {Knorr}}, \ and\ \bibinfo {author} {\bibfnamefont {Ermin}\ \bibnamefont
  {Malic}},\ }\bibfield  {title} {\enquote {\bibinfo {title} {Dark and bright
  exciton formation, thermalization, and photoluminescence in monolayer
  transition metal dichalcogenides},}\ }\href {\doibase
  10.1088/2053-1583/aabea3} {\bibfield  {journal} {\bibinfo  {journal} {2D
  Materials}\ }\textbf {\bibinfo {volume} {5}},\ \bibinfo {pages} {035017}
  (\bibinfo {year} {2018})}\BibitemShut {NoStop}%
\bibitem [{\citenamefont {{Paradisanos}}\ \emph {et~al.}(2020)\citenamefont
  {{Paradisanos}}, \citenamefont {{Wang}}, \citenamefont {{Alexeev}},
  \citenamefont {{Cadore}}, \citenamefont {{Marie}}, \citenamefont {{Ferrari}},
  \citenamefont {{Glazov}},\ and\ \citenamefont
  {{Urbaszek}}}]{2020arXiv200705369P}%
  \BibitemOpen
  \bibfield  {author} {\bibinfo {author} {\bibfnamefont {Ioannis}\ \bibnamefont
  {{Paradisanos}}}, \bibinfo {author} {\bibfnamefont {Gang}\ \bibnamefont
  {{Wang}}}, \bibinfo {author} {\bibfnamefont {Evgeny~M.}\ \bibnamefont
  {{Alexeev}}}, \bibinfo {author} {\bibfnamefont {Alisson~R.}\ \bibnamefont
  {{Cadore}}}, \bibinfo {author} {\bibfnamefont {Xavier}\ \bibnamefont
  {{Marie}}}, \bibinfo {author} {\bibfnamefont {Andrea~C.}\ \bibnamefont
  {{Ferrari}}}, \bibinfo {author} {\bibfnamefont {Mikhail~M.}\ \bibnamefont
  {{Glazov}}}, \ and\ \bibinfo {author} {\bibfnamefont {Bernhard}\ \bibnamefont
  {{Urbaszek}}},\ }\bibfield  {title} {\enquote {\bibinfo {title} {{Efficient
  phonon cascades in hot photoluminescence of WSe$_2$ monolayers}},}\
  }\href@noop {} {\bibfield  {journal} \Eprint
  {http://arxiv.org/abs/2007.05369} {arXiv:2007.05369 (2020)}}
  \BibitemShut {NoStop}%
\bibitem [{\citenamefont {Glazov}(2020)}]{PhysRevLett.124.166802}%
  \BibitemOpen
  \bibfield  {author} {\bibinfo {author} {\bibfnamefont {M.~M.}\ \bibnamefont
  {Glazov}},\ }\bibfield  {title} {\enquote {\bibinfo {title} {Quantum
  interference effect on exciton transport in monolayer semiconductors},}\
  }\href {\doibase 10.1103/PhysRevLett.124.166802} {\bibfield  {journal}
  {\bibinfo  {journal} {Phys. Rev. Lett.}\ }\textbf {\bibinfo {volume} {124}},\
  \bibinfo {pages} {166802} (\bibinfo {year} {2020})}\BibitemShut {NoStop}%
\bibitem [{\citenamefont {Glazov}(2007)}]{glazov2007}%
  \BibitemOpen
  \bibfield  {author} {\bibinfo {author} {\bibfnamefont {M.~M.}\ \bibnamefont
  {Glazov}},\ }\bibfield  {title} {\enquote {\bibinfo {title} {Effect of
  structure anisotropy on low temperature spin dynamics in quantum wells},}\
  }\href@noop {} {\bibfield  {journal} {\bibinfo  {journal} {Solid State
  Commun.}\ }\textbf {\bibinfo {volume} {142}},\ \bibinfo {pages} {531}
  (\bibinfo {year} {2007})}\BibitemShut {NoStop}%
\bibitem [{\citenamefont {Nickolaus}\ \emph {et~al.}(1998)\citenamefont
  {Nickolaus}, \citenamefont {W\"unsche},\ and\ \citenamefont
  {Henneberger}}]{PhysRevLett.81.2586}%
  \BibitemOpen
  \bibfield  {author} {\bibinfo {author} {\bibfnamefont {H.}~\bibnamefont
  {Nickolaus}}, \bibinfo {author} {\bibfnamefont {H.-J.}\ \bibnamefont
  {W\"unsche}}, \ and\ \bibinfo {author} {\bibfnamefont {F.}~\bibnamefont
  {Henneberger}},\ }\bibfield  {title} {\enquote {\bibinfo {title} {Exciton
  spin relaxation in semiconductor quantum wells: The role of disorder},}\
  }\href {\doibase 10.1103/PhysRevLett.81.2586} {\bibfield  {journal} {\bibinfo
   {journal} {Phys. Rev. Lett.}\ }\textbf {\bibinfo {volume} {81}},\ \bibinfo
  {pages} {2586--2589} (\bibinfo {year} {1998})}\BibitemShut {NoStop}%
\bibitem [{\citenamefont {Gridnev}(2001)}]{gridnev01}%
  \BibitemOpen
  \bibfield  {author} {\bibinfo {author} {\bibfnamefont {V.~N.}\ \bibnamefont
  {Gridnev}},\ }\bibfield  {title} {\enquote {\bibinfo {title} {Theory of
  Faraday rotation beats in quantum wells with large spin splitting},}\
  }\href@noop {} {\bibfield  {journal} {\bibinfo  {journal} {JETP Letters}\
  }\textbf {\bibinfo {volume} {74}},\ \bibinfo {pages} {380} (\bibinfo {year}
  {2001})}\BibitemShut {NoStop}%
\bibitem [{\citenamefont {Zhu}\ \emph {et~al.}(2014)\citenamefont {Zhu},
  \citenamefont {Zhang}, \citenamefont {Glazov}, \citenamefont {Urbaszek},
  \citenamefont {Amand}, \citenamefont {Ji}, \citenamefont {Liu},\ and\
  \citenamefont {Marie}}]{PhysRevB.90.161302}%
  \BibitemOpen
  \bibfield  {author} {\bibinfo {author} {\bibfnamefont {C.~R.}\ \bibnamefont
  {Zhu}}, \bibinfo {author} {\bibfnamefont {K.}~\bibnamefont {Zhang}}, \bibinfo
  {author} {\bibfnamefont {M.}~\bibnamefont {Glazov}}, \bibinfo {author}
  {\bibfnamefont {B.}~\bibnamefont {Urbaszek}}, \bibinfo {author}
  {\bibfnamefont {T.}~\bibnamefont {Amand}}, \bibinfo {author} {\bibfnamefont
  {Z.~W.}\ \bibnamefont {Ji}}, \bibinfo {author} {\bibfnamefont {B.~L.}\
  \bibnamefont {Liu}}, \ and\ \bibinfo {author} {\bibfnamefont
  {X.}~\bibnamefont {Marie}},\ }\bibfield  {title} {\enquote {\bibinfo {title}
  {Exciton valley dynamics probed by Kerr rotation in ${\mathrm{WSe}}_{2}$
  monolayers},}\ }\href {\doibase 10.1103/PhysRevB.90.161302} {\bibfield
  {journal} {\bibinfo  {journal} {Phys. Rev. B}\ }\textbf {\bibinfo {volume}
  {90}},\ \bibinfo {pages} {161302} (\bibinfo {year} {2014})}\BibitemShut
  {NoStop}%
\bibitem{brand02}
M.~A. Brand, A.~Malinowski, O.~Z. Karimov, P.~A. Marsden, R.~T. Harley, A.~J.
  Shields, D.~Sanvitto, D.~A. Ritchie, and M.~Y. Simmons, ``Precession and motional slowing of spin evolution in a high mobility
  two-dimensional electron gas,'' Phys. Rev. Lett. {\bf 89}, 236601 (2002).

\end{thebibliography}
\end{document}